\documentclass[aps,prl,twocolumn,superscriptaddress,longbibliography]{revtex4-1}

\usepackage{graphicx}
\usepackage{bm}
\usepackage{hyperref}
\usepackage{amsmath}
\usepackage{amssymb}
\usepackage[table]{xcolor}
\usepackage{array}
\usepackage{multirow}
\usepackage[caption=false]{subfig}
\usepackage{physics}
\usepackage{amsthm}
\usepackage{bbm}
\usepackage{multirow}
\usepackage[colorinlistoftodos]{todonotes}
\usepackage[scr=boondoxo,frak=boondox,scrscaled=1.05]{mathalfa}
\usepackage{amsmath}

\def\ket#1{|#1\rangle}
\def\braket#1#2{\langle#1|#2\rangle}

\def\up{\uparrow}

\usepackage{ulem}
\begin{document}
\title{Non-Abelian spin Hall insulator}
\author{Ahmed Abouelkomsan}
\email{ahmed95@mit.edu}
\affiliation{Department of Physics, Massachusetts Institute of Technology, Cambridge, Massachusetts 02139, USA}
\author{Liang Fu}
\email{liangfu@mit.edu}
\affiliation{Department of Physics, Massachusetts Institute of Technology, Cambridge, Massachusetts 02139, USA}

\begin{abstract}
    Motivated by a recent experiment reporting the fractional quantum spin Hall effect in twisted ${\rm MoTe}_2$, we investigate microscopically the prospects of realizing exotic topologically ordered states beyond conventional quantum Hall physics. We show that a \textit{non-Abelian} spin Hall insulator, a state of two copies of the non-Abelian Moore-Read state, can be stabilized at half filling of time-reversal conjugate Chern bands. We elucidate that the existence of this phase relies on the reduction of opposite-spin interactions at short distances to overcome the Ising ferromagnetism. Moreover, we demonstrate that band mixing provides a generic mechanism for this reduction to be achieved. Quite remarkably, we find that a renormalization of opposite-spin interactions at short distances as small as 15 \% of the moir\'e period is sufficient for a \textit{direct} transition to a completely spin unpolarized phase which supports the non-Abelian spin Hall insulator. Furthermore, we show that the non-Abelian spin Hall insulator can either break time-reversal symmetry or preserve it depending on the underlying topological order. 
\end{abstract}

\maketitle

\section{I. Introduction}

Electron fractionalization is a central theme of modern condensed matter physics. About four decades ago, the study of the fractional quantum Hall effect of two-dimensional electron systems under high magnetic field revealed the existence of quasiparticles in a partially filled Landau level  which carry fractional charge and obey fractional statistics \cite{tsui1982two,laughlin_anomalous_1983}. Recently, fractional quantum anomalous Hall states were discovered in two-dimensional semiconductor and graphene moir\'e systems at zero magnetic field \cite{zeng2023thermodynamic,cai2023signatures,park2023observation,xu_observation_2023, Lu2024Feb}. In both cases, the emergence of fractional quasiparticle requires breaking time-reversal symmetry due to external magnetic field or spontaneous ferromagnetism \cite{li2021spontaneous, crepel2023anomalous}, and is marked by the fractionally quantized Hall conductance. On the other hand, as the vast majority of quantum materials are nonmagnetic,   
there is growing interest in the possibility of electron fractionalization with time reversal symmetry  \cite{levin_fractional_2009, neupert_fractional_2011,freedman_class_2004, repellin_mathbbz_2_2014,bernevigQuantumSpinHall2006}. 

Topological quantum materials with time-reversed pairs of Chern bands \cite{kane2005quantum}, such as twisted homobilayer MoTe$_2$ and WSe$_2$ \cite{wu_topological_2019, devakul2021magic},  
provide a promising venue for fractional electron states. While most studies have focused on fractional Chern insulators in which spontaneously spin-polarized electrons populate a single Chern band \cite{cai2023signatures,park2023observation,zeng2023thermodynamic,xu_observation_2023}, a recent transport experiment \cite{kang_evidence_2024} has reported an incompressible state in twisted $\text{MoTe}_2$ at odd-integer hole filling $\nu= 3$, which has vanishing Hall conductivity accompanied by fractional edge conductance $\frac{3}{2}e^2/\hbar$ \textcolor{black}{per spin}. 
This sharply contrasts with the quantum anomalous Hall (QAH) state at $\nu= 1$ with complete filling of a single Chern band \cite{devakul2021magic, fouttyMappingTwisttunedMultiband2024}, or the (double) quantum spin Hall states at $\nu=2$ or $4$ with complete filling of time-reversed pairs of Chern bands \cite{devakul2021magic, wu_topological_2019, kang2024observationWSe2}.   

The observed $\nu=3$ state has attracted great interest. A number of time-reversal-invariant and time-reversal-breaking phases with different fractional excitations has been proposed on phenomenological grounds \cite{may2024theory, jian2024minimal,villadiego2024halperin,zhang2024non,chou2024composite}. However, no microscopic calculation has been performed to shine light on the likely ground states of twisted semiconductor bilayers at $\nu=3$.

In this work, we identify two competing strong-coupling phases at $\nu = 3$, a quantum anomalous Hall insulator and a non-Abelian spin Hall insulator. We demonstrate that the competition between the two phases is governed by the  behavior of electron-electron interactions at short distances which we systematically investigate. In addition, we establish band mixing as a microscopic mechanism for controlling the effective interaction at short distance and therefore the competition between the two phases.

Our results are obtained through a  microscopic model study of interacting electron systems featuring time-reversed pairs of narrow Chern bands with $C_\uparrow=-C_\downarrow=1$ at odd-integer band fillings $\nu\equiv \nu_\uparrow + \nu_\downarrow = 1 \mod 2$.   
In the flat band limit, if the gap to higher bands is large so that band mixing can be neglected,  Coulomb interaction drives the system into a fully spin-polarized QAH state. 
However, we find that a reduction of the short-range part of Coulomb repulsion between two opposite-spin electrons  can result in a spin unpolarized state with both $C_\uparrow=1$ and $C_\downarrow=-1$ bands at equal half-integer filling.


We further show that if the underlying Chern band wavefunctions mimics that of the $n=1$ Landau level \cite{reddy2024non}, the spin unpolarized ground state is a non-Abelian spin Hall insulator, which is adiabatically connected to the product of two non-abelian Moore-Read states  \cite{Moore1991Aug,Read2000Apr} in conjugate Chern bands. This state has vanishing electrical Hall conductivity, quantized spin Hall conductivity $\sigma^{sh} = \frac{3}{2}$, and supports both charge $e/4$ non-Abelian Ising anyons and $e/2$ Abelian anyons in the bulk. Interestingly, two types of non-Abelian spin Hall insulators are identified: a time-reversal-invariant phase (a non-Abelian topological insulator), and a time-reversal-breaking phase that exhibits quantized thermal Hall conductance $\kappa_H= \pm 2$. 
Remarkably, our non-Abelian state only requires a small renormalization of Coulomb interaction at short distance (which can be as small as $15\%$ of the moir\'e  period), and can  already emerge at a moderate amount of band mixing. 


Our work is organized as follows. We start by introducing a minimal model that captures the essential feature of conjugate Chern bands in twisted TMD bilayers (section II). Since spin $\uparrow$ and $\downarrow$ particles belong to conjugate Chern bands of opposite chirality, our system lacks spin $SU(2)$ symmetry and its energy spectrum depends crucially on the total spin (section III).   

Next, we show rigorously that in the absence of band mixing, the ground state of our model at odd integer fillings is necessarily fully spin polarized if the bare interaction is spin independent and repulsive (section IV). However,  when opposite-spin repulsion is reduced at short distance, we find by analytical calculations that the ferromagnetic state can become unstable to a single spin flip (section V). 

This motivates us to study the ground state phase diagram by many-body exact diagonalization calculation (section VI). Remarkably, as the opposite-spin repulsion is reduced, the fully polarized state transitions directly into an unpolarized state. The latter has an approximate ground state degeneracy of $6\times6$ and $2\times 2$ for even and odd number of particles respectively, as expected for the product of two Moore-Read states in opposite magnetic fields. 
Two types of ``doubled'' Moore-Read phases are discussed, which are time-reversal-invariant and spontaneous time-reversal-breaking respectively (section VII).

Moreover, we show quantitatively that the reduction of opposite-spin interaction at short distance naturally arises  from the band mixing effect (section VIII). 

Finally, we discuss the relevance of our finding for the observed $\nu=3$ state in twisted MoTe$_2$ (section IX) and outline future directions (section X). 

\section{II. Model}  
Our work is motivated by the physics of twisted transition metal dichalcogenide homobilayers ($t$TMDs) such as $t$MoTe$_2$, described by the following Hamiltonian: 
\begin{equation}
    \begin{gathered}
            H = \int d^2r \psi_{\sigma}^{\dagger}(\mathbf{r}) \mathcal{H}_{\sigma}\psi_{\sigma}(\mathbf{r})+ 
           H_{\rm int} 
\end{gathered}
\end{equation}
where $\psi_{\sigma}^{\dagger}(\mathbf{r})$ with $\sigma=\pm$ creates a  spin $s_z=\uparrow$ or $\downarrow$ particle (which corresponds to a hole excitation in the valence band). $\mathcal{H}$ describes the non-interacting band structure of a single hole. Due to strong Ising spin-orbit coupling, our system only has spin $U(1)$ symmetry, and spin $\uparrow$ and $\downarrow$ bands form time-reversed pairs, satisfying $\mathcal{H}_{+}=\mathcal{H}^*_-$.  
The second term $H_{\rm int}$ describes the interaction between holes which we will specify later. 

The key feature of $t$TMDs is time-reversed pairs of topological moir\'e bands, which carry a spin-dependent Chern number $C_\uparrow=-C_\downarrow=1$ \cite{wu_topological_2019}. These Chern bands are formed by a real-space topological texture of “Zeeman” field acting on the layer pseudospin degree of freedom. 
When the Zeeman field is large, the pseudospin is locally aligned to the Zeeman field, giving rise to an effective Hamiltonian description of the low-energy band structure \cite{paul_magic2022, morales2023magic}, 
\begin{equation}
\label{eq:effectiveHam}
    \mathcal{H}_{\sigma} = \frac{(\mathbf{p} + \sigma e \mathbf{A}(\mathbf{r}))^2}{2m} + V(\mathbf{r}).
\end{equation}
Here, $\mathbf{A}(\mathbf{r})$ is a $U(1)$ gauge field that represents an emergent magnetic field $B(\mathbf{r})$ with one flux quanta per moir\'e unit cell, while $V(\mathbf{r})$ is a scalar potential. Due to time-reversal symmetry, opposite spins feel opposite magnetic fields and identical scalar potential. Both $B(\mathbf{r})$ and $V(\mathbf{r})$ are periodically varying with the periodicity of the moir\'e superlattice.  
The Hamiltonian \eqref{eq:effectiveHam}, which describes a charged particle subject to a periodic magnetic field and a periodic potential, defines the adiabatic model of twisted TMDs.

It is instructive to first study the simplified version of the Hamiltonian \eqref{eq:effectiveHam}, where we only keep the uniform component of the emergent magnetic field, neglecting other Fourier components of $B(\mathbf{r})$ and $V(\mathbf{r})$ at nonzero reciprocal lattice vectors. In this approximation---which turns out to be accurate near a magic twist angle \cite{morales2023magic, reddy2024non},    
the moir\'e bands of our system are simply conjugate pairs of Landau levels with opposite spins (LLs), produced by equal and opposite magnetic fields and related by time-reversal symmetry. We label the time-reversed pair of LLs as ($n$LL, $n\overline{\rm LL}$) with $n=0, 1, ...$ the Landau level index.   

When the band gap is large enough that the effect of band mixing can be neglected, at total band filling $\nu=2n+1$, the lowest $n$  LL pairs are completely filled and inert, while interacting fermions in the next LL pair (corresponding to LL index $n$) at effective filling $\nu_\uparrow + \nu_\downarrow=1$ is described by the Hamiltonian  
\begin{equation}
\label{eq:Ham_int}
    \begin{split}
               \tilde{H} = \frac{1}{2S_M} \sum_{\mathbf{q} \sigma} V_n(\mathbf{q}):\rho_{\sigma}(\mathbf{q}) \rho_{\sigma}(-\mathbf{q}): 
               \\ \> \> \> \> \> \> \> +  U_n (\mathbf{q}):\rho_{\sigma}(\mathbf{q}) \rho_{-\sigma}(-\mathbf{q}):
    \end{split}
\end{equation}
where $\rho_{\sigma}(\mathbf{q}) $ is the projected density operator for spin $\sigma$ given by $ \rho_{\sigma}(\mathbf{q}) = \sum_{\mathbf{k}} \langle \mathbf{k}+\mathbf{q}|e^{i\mathbf{q} \cdot \mathbf{R}^{\sigma}}| \mathbf{k} \rangle c^{\dagger}_{\mathbf{k}+\mathbf{q}\sigma} c_{\mathbf{k}\sigma}$ with $\mathbf{R}^{\sigma}$ the guiding center coordinate \cite{supp}. $S_M$ is the area of the moir\'e system and $: :$ denotes normal ordering.

The density operators of equal spin satisfy the {\it spin-dependent} GMP algebra \cite{girvinMagnetorotonTheoryCollective1986},
\begin{eqnarray}
[\rho_{\sigma}(\mathbf{q}_1),\rho_{\sigma}(\mathbf{q}_2)] &=& 2 i \sigma \sin( l^2_B \mathbf{q}_1 \wedge \mathbf{q}_2)  \rho_{\sigma}(\mathbf{q}_1 + \mathbf{q}_2)
\end{eqnarray} 
where $l_B$ is the magnetic length set by one emergent flux quanta per moir\'e unit cell, $2 \pi l^2_B = \frac{\sqrt{3}}{2} a_M^2$ with $a_M$ the moir\'e period. It is important to note that 
the commutators for spin-$\uparrow$ and $\downarrow$ take opposite signs, reflecting the opposite chirality of the corresponding LLs. The density operators of different spins commute: 
\begin{eqnarray}
[\rho_{+}(\mathbf{q}_1),\rho_{-}(\mathbf{q}_2)] = 0.
\end{eqnarray}
The band-projected Hamiltonian \eqref{eq:Ham_int}  defines a concrete model we shall study below.

In the absence of moir\'e band mixing, $V_n(\mathbf{q})$ and $U_n(\mathbf{q})$ are obtained from the interaction between two equal- and opposite-spin particles projected onto ($n$LL, $n\overline{\rm LL}$). For the bare Coulomb interaction $V(\mathbf{r})=e^2/(\epsilon \> \mathbf{r})$, we find 
\begin{eqnarray}
V_n(\mathbf{q}) = \left( \frac{2 \pi e^2}{\epsilon |\mathbf{q}|}  \right) e^{-|\mathbf{q}|^2 l^2_B/2} [L_n(|\mathbf{q}|^2 l_B^2/2)]^2   \label{V}  
\end{eqnarray}
where $L_n$ is the $n$-th Laguerre polynomial, and $U_n(\mathbf{q}) =V_n(\mathbf{q})$ due to the time reversal relation between conjugate LLs. 
While the long-range part of electron-electron interaction described by Coulomb's law is indeed spin independent, the interaction at short distance is generally {\it spin-dependent}, especially considering the strong atomic spin-orbit coupling in TMD. Moreover, as we shall show explicitly later, moir\'e band mixing will strongly reduce the opposite-spin interaction $U_n$ compared to the equal-spin interaction $V_n$.

Keeping these considerations in mind, in this work we consider projected interactions on a pair of conjugate Chern bands that are spin dependent. In particular, we find it useful to consider interactions in the $n$-th pair of conjugate LLs of the form: 
\begin{eqnarray}
\label{eq:oppositespin_int}
U_n(\mathbf{q}) = e^{-d |\mathbf{q}|} V_n(\mathbf{q})  \label{U}   
\end{eqnarray} 
where $d$ is a spin anisotropy parameter. It is important to note that the introduction of $d$ mainly affects the short distance behavior of the \textit{projected} interaction between opposite-spin particles, while at long distance ($\mathbf{q} \rightarrow 0$), $U_n$ and $V_n$ converge to the same value as physically required. 

The interactions defined in \eqref{V} and \eqref{U} have a physically transparent meaning. At small $d$,  $U_n$ can be rewritten as $U_n=V_n  + \delta U_n$ with the correction $\delta U_n$ given by  
\begin{eqnarray}
\delta U_n &=&  - (1 - e^{-d |\mathbf{q}|}) V_n(\mathbf{q})   \nonumber \\
&\approx & -\left( \frac{2 \pi e^2 d}{\epsilon} \right)e^{-|\mathbf{q}|^2 l^2_B/2} [L_n(|\mathbf{q}|^2 l_B^2/2)]^2.   
\end{eqnarray}
This corresponds to a delta-function attraction between opposite spins projected onto the $n$-th conjugate LLs, added to the spin-independent Coulomb interaction. In other words, the role of  small $d$ is to reduce the electron-electron interaction between two opposite-spin particles at short distances from the Coulomb case while the overall interaction remains repulsive at all distances. In the opposite limit $d\rightarrow \infty$, spin-$\uparrow$ and $\downarrow$ particles are essentially decoupled, which will greatly aid our analysis below.

\section{III. Two-Body Energy Spectrum}
The interaction $V_n$ between two particles of equal spin in the $n$-th LL is familiar in the quantum Hall literature.  Assuming full rotational invariance, it can be naturally decomposed into the standard Haldane pseudopotentials \cite{haldane1983fractional} of the $n$-th LL denoted as $V_n^m$ with $m$ the relative angular momentum, 
\begin{equation}
\label{eq:Haldane_pseudo}
    V_n^m = \int \dfrac{d^2\mathbf{q}}{(2 \pi)^2} V(\mathbf{q}) [L_n(|\mathbf{q}|^2 l^2_B/2)]^2 L_m(|\mathbf{q}|^2 l^2_B) e^{-|\mathbf{q}|^2 l^2_B}
\end{equation}

On the other hand, $U_n$ describes the interaction between two particles of opposite spin in {\it conjugate} LLs, which doesn't allow for a similar decomposition. Nonetheless, the two-body problem of $U_n$ can still be solved \cite{stefanidis_excitonic_2020} due to the algebra of guiding center relative coordinates, $\mathbf{R}^{\rm rel} = \mathbf{R}^{\uparrow}_{1} - \mathbf{R}^{\downarrow}_{2}$ which obeys 
\begin{eqnarray}
[\mathbf{R}^{\rm rel}_a,\mathbf{R}^{\rm rel}_b] = 0
\end{eqnarray} 
for $a,b = x,y$ \cite{supp}. The two-body energy spectrum can then be labelled by  $\mathbf{R}^{\rm rel}$ and are given by \begin{equation}
\label{eq:V0intra}
    U_n({\mathbf{R}^{\rm rel}}) = \int \frac{d^2\mathbf{q}}{(2\pi)^2} U_n(\mathbf{q})  e^{i \mathbf{q} \cdot \mathbf{R}^{\rm rel}}
\end{equation}
The highest energy state corresponds to a bound state of two particles with a vanishing separation $\mathbf{R}^{\rm rel} = 0$, which has the largest repulsion energy $U_n( 0)$.

In addition, since two particles of opposite spin interacting by $U_n$ feel opposite magnetic fields, they drift parallel to each other with total momentum perpendicular to their separation, $\mathbf{K} = - \hat{z} \cross \mathbf{R}^{\rm rel}/ l^2_B $ in an analogous manner to neutral excitons in Landau levels \cite{gor1968contribution,kallin_excitations_1984}, leading to a dispersive band of two-body excitations with total spin $S_z=0$.  

We now focus on the $n = 1 $ LL,  the equal-spin pseudopotentials $V_1^m$ read
\begin{equation}
\label{eq:Haldane_1LL}
   V_1^m = \frac{e^2}{\epsilon l_B} \dfrac{\Gamma(m+1/2)}{2 m !} \dfrac{(m-3/8)(m-11/8)}{(m-1/2)(m-3/2)}
\end{equation} and the largest repulsion energy for opposite-spin interactions is given by
\begin{widetext}
    \begin{equation}
    \label{eq:U10}
U_1(0)  =   \dfrac{e^2}{\epsilon l_B}  \dfrac{1}{8}\big[-2\big(\frac{d}{l_B} + \frac{d^3}{l_B^3}\big)+ \big(3 + 2 \frac{d^2}{l_B^2} + \frac{d^4}{l_B^4}\big) e^{d^2/ 2l_B^2} \sqrt{2 \pi} {\rm erfc}(\dfrac{d}{\sqrt{2}l_B})\big]
\end{equation}
\end{widetext}
For later discussions, we define $\eta = U_1(0)/\frac{3} {4} \frac{e^2}{\epsilon l_B}\sqrt{\frac{\pi}{2}}$ to be the ratio of the largest opposite-spin repulsion energy at any $d$ to its value when $d =0 $.

\begin{figure}[t!]
    \centering
    \includegraphics[width =\linewidth]{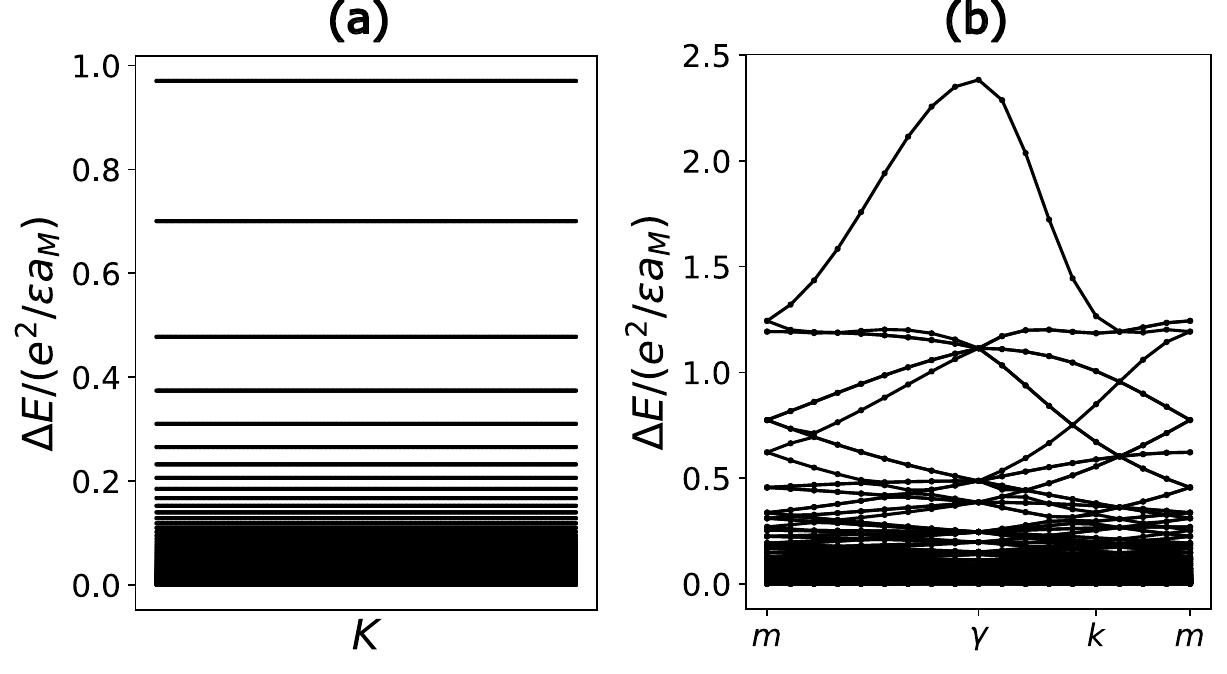}
    \caption{Two-body spectrum of the Hamiltonian \eqref{eq:Ham_int} for $V_n = U_n$ evaluated in the $n = 1$ LL when the spin of the two particles is equal (a) or opposite (b). $m-\gamma-k-m$ is a symmetry cut in the Brillouin zone \textcolor{black}{of the center of mass momentum of the two particles.}. The full two-body spectrum is obtained from numerical exact diagonalization }
    \label{fig:twobodyspectrum}
\end{figure}

As shown in Fig. \ref{fig:twobodyspectrum},  the two-particle energy spectrum in the $n = 1$ LL for total spin $S_z=\pm 1$ and $0$ at $d=0$ (or $U_n=V_n$)  are markedly different. The absence of spin $SU(2)$ symmetry results from the fact that equal-spin particles belong to the same Landau level, whereas opposite-spin particles belong to conjugate Landau levels of opposite chirality. Given the strong dependence of the energy spectrum on the total spin $S_z$, in order to understand our system at finite density it is essential to first determine its ground state spin polarization.



\section{IV. Ferromagnetism}
At the filling $\nu=1$ of the $n$-th conjugate LL pair, the fully spin polarized state with one of the two LLs completely filled is always an energy eigenstate of the band-projected Hamiltonian \eqref{eq:Ham_int}, because there is no matrix element connecting it to any other state.  However, the fully polarized state may or may not be the ground state, depending on the competition between equal- and opposite-spin interactions $V_n$ and $U_n$. Obviously, the fully polarized state completely avoids the opposite-spin interaction at the expense of costing equal-spin interaction energy.

We now prove an exact result for ferromagnetism. As long as the bare interaction is spin independent and repulsive, the ground state of our system in the absence of band mixing is the fully polarized QAH state, which spontaneously breaks the $Z_2$  symmetry.  
To show this, we note that a bare interaction that is spin-independent and repulsive yields, upon projection onto the conjugate LL pair, $U_n = V_n \geq 0$ for all $\mathbf{q}$. Then, the projected Hamiltonian can be expressed in terms of the projected {\it total density} operator $\rho \equiv \rho_+ + \rho_-$ as follows:  
\begin{equation}
\label{eq:Ham}
               \tilde{H}= 
               \frac{1}{2S_M} \sum_{\mathbf{q} \neq 0} V_n(\mathbf{q}) \rho(\mathbf{q}) \rho(-\mathbf{q}).  
\end{equation}
Here we have dropped the $\mathbf{q}=0$ term and a one-body term from normal ordering: both terms in the LL problem only depend on the total number of particles which is fixed at $\nu=1$.  Importantly, Eq.\eqref{eq:Ham} is manifestly positive semi-definite. Therefore,  the fully polarized state with a single LL completely filled is annihilated by the projected density operator $\rho(\mathbf{q})$ for any $\mathbf{q} \neq 0$, and is necessarily a ground state of the band projected Hamiltonian.

Our exact result on full spin polarization in conjugate LL systems is consistent with a previous study of valley-degenerate LL systems with valley-dependent LL wavefunctions (which found the fully polarized ground state state within the Hartree-Fock approximation) \cite{sodemann_quantum_2017}, as well as related results on ferromagnetism in certain flat band systems \cite{bultinck_ground_2020,bultinck_mechanism_2020,lian_twisted_2021,repellin_ferromagnetism_2020}.
More importantly, our proof offers important insight into the necessary condition for the breakdown of ferromagnetism, that is, only if the repulsive interaction in the conjugate LLs is spin-dependent ($V_n \neq U_n) $.

\section{V. Magnon Instability}

\begin{figure}[t!]
    \centering
    \includegraphics[width =0.9 \linewidth]{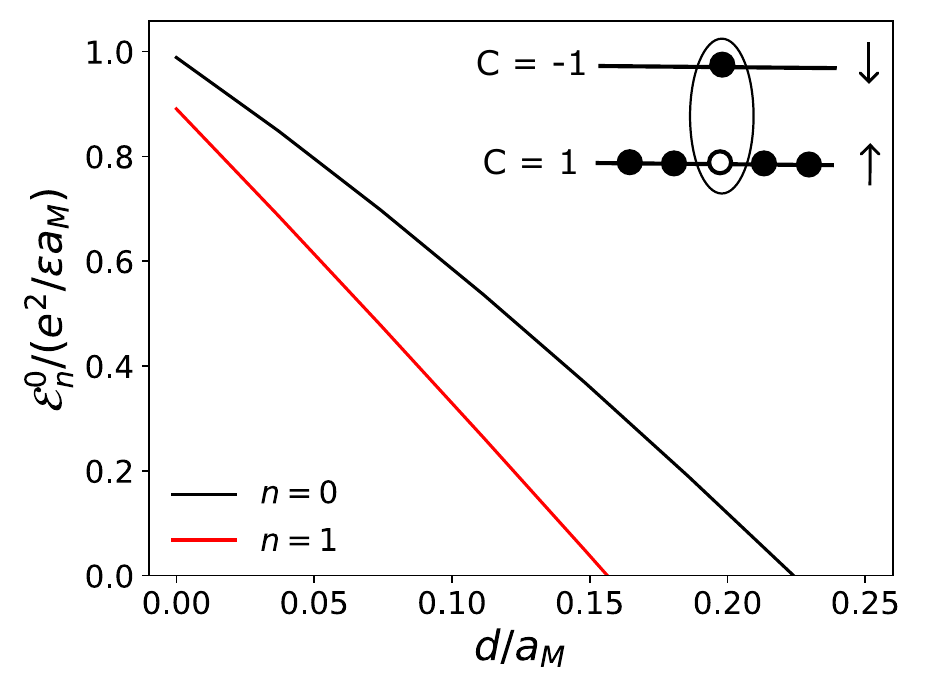}
    \caption{The lowest magnon energy $\mathcal{E}_n^0$  (equation \eqref{eq:magnon01}) as a function of $d/aM$ calculated for the $n = 0$ and $ n= 1$ Landau levels. }
    \label{fig:magnon}
\end{figure}

To investigate the stability of ferromagnetism in our system under spin-dependent interactions, we study the energetics of the underlying single spin-flip excitation (magnon). Due to the breaking of spin $SU(2)$ symmetry by conjugate LLs,  
the magnon spectrum  is expected to be gapped when the ground state is fully polarized. On the other hand, a negative magnon energy relative to the fully polarized state indicates that the true ground state is non-fully polarized.  In this section, we analytically calculate the magnon energy in our model as a function of the spin anisotropy parameter $d$, and find it vanishes at a small $d$, which is about $15\%$ of the moire period for the $n=1$ conjugate LL (see Fig. \ref{fig:magnon}). Thus, a small reduction of opposite-spin repulsion at short distance suffices to drive our system out of the fully polarized regime.  

It is important to note the magnon in our system is a bound state of a particle and a hole in {\it conjugate} LLs and therefore experiences an effective magnetic field \cite{kwan_exciton_2021,kwan_excitonic_2022,stefanidis_excitonic_2020}. As a result, the magnon energy spectrum consists of flat bands, i.e., magnon Landau levels. This should be contrasted with the magnon in quantum Hall systems which feels zero magnetic field and has a dispersive energy-momentum relation.   


For the analysis below, we find it convenient to perform a particle-hole (PH) transformation \cite{abouelkomsan2020particle} of the Hamiltonian \eqref{eq:Ham_int} on one of the spins: $c^{\dagger}_{\mathbf{k} \downarrow} \rightarrow d_{\mathbf{k}\downarrow}$. The transformed Hamiltonian then reads: 
\begin{multline}
         \tilde{H} = \frac{1}{2S_M} \sum_{\mathbf{q}} V_n(\mathbf{q}) \big[\!:\! \rho_{\uparrow}(\mathbf{q}) \rho_{\uparrow}(-\mathbf{q})\!: + :\!\overline{\rho}_{\downarrow}(\mathbf{q}) \overline{\rho}_{\downarrow}(-\mathbf{q})\!:\! \big]
               \\ -  \frac{1}{S_M} \sum_{\mathbf{q}} U_n (\mathbf{q}): \! \rho_{\uparrow}(\mathbf{q}) \overline{\rho}_{\downarrow}(-\mathbf{q})\!: \\ + \sum_{\mathbf{k} \in {\rm BZ}} \big[E^{\rm HF}_{\downarrow}(\mathbf{k}) \hat{N}_{\downarrow}(\mathbf{k}) + E^{\rm H}_{\uparrow}(\mathbf{k}) \hat{N}_{\uparrow}(\mathbf{k})\big]
\label{eq:PH_Hamint}
\end{multline}
where $\overline{\rho}_{\downarrow}(\mathbf{q}) \sum_{\mathbf{k}} \langle \mathbf{k} + \mathbf{q} |e^{i\mathbf{q}\cdot \mathbf{R}_{\downarrow}}|\mathbf{k}\rangle d^{\dagger}_{\mathbf{k} \downarrow} d_{\mathbf{k} + \mathbf{q},\downarrow}$, $\hat{N}_{\sigma}(\mathbf{k})$ is the number operator for spin $\sigma$ and 
\begin{eqnarray}
                E^{\rm HF}_{\downarrow}(\mathbf{k)} &=& \frac{1}{S_M} \sum_{\mathbf{k}' \in {\rm BZ}} V_n(\mathbf{k}' \mathbf{k} \mathbf{k}' \mathbf{k}) -  V_n(\mathbf{k'}\mathbf{k}\mathbf{k}\mathbf{k}') \nonumber \\
    E^{\rm H}_{\uparrow}(\mathbf{k}) &=& \frac{1}{S_M} \sum_{\mathbf{k'} \in {\rm BZ}} U_n(\mathbf{k}'\mathbf{k} \mathbf{k} \mathbf{k}')
\end{eqnarray}
represent the Hartree-Fock energy of a hole in a completely filled spin-$\downarrow$ LL and the Hartree energy it exerts on a spin-$\uparrow$ particle, respectively.  
Here $V_n(\mathbf{k}_1 \mathbf{k}_2\mathbf{k}_3\mathbf{k}_4) = \langle \mathbf{k}_1\sigma, \mathbf{k}_2\sigma | V_n  | \mathbf{k}_3 \sigma , \mathbf{k}_4 \sigma \rangle$ are the matrix elements of same-spin interactions and similarly for $U_n(\mathbf{k}_1 \mathbf{k}_2 \mathbf{k}_3 \mathbf{k}_4) = \langle \mathbf{k}_1\!\! \uparrow, \mathbf{k}_2\!\! \downarrow \!\!|U_n| \mathbf{k}_4 \! \!\uparrow, \mathbf{k}_3\!\! \downarrow \rangle$. Due to magnetic translation symmetry of Landau levels, both $E^{\rm HF}_{\downarrow}(\mathbf{k})$ and $E^{\rm H}_{\uparrow}(\mathbf{k})$ are independent of $\mathbf{k}$. 

The above PH transformation on one of the spins maps our Hamiltonian of two time-reversal conjugate LLs to one describing two LLs in the \textit{same} magnetic field as evident from the conjugated matrix elements in the the density operator $\overline{\rho}_{\downarrow}(\mathbf{q})$. However, the opposite-spin interactions become \textit{attractive} as indicated from minus sign in front of $U_n$ in equation \eqref{eq:PH_Hamint}. The magnon binding energies simply become the Haldane pseudopotentials of $U_n(\mathbf{q})$. Relative to the fully polarized state, the magnon energy spectrum in the $n$-th LL takes the general form
\begin{equation}
\label{eq:magnon}
    \mathcal{E}^m_{n} = E^{\rm HF}_{\downarrow} + E^{\rm H}_{\uparrow} - U_n^m 
\end{equation}
with the Haldane pseudopotentials $U_n^m$ defined similar to equation \eqref{eq:Haldane_pseudo} but with $U(\mathbf{q})$ instead.  
Consider first the case where equal spin and opposite spin interactions are the same ($V_n = U_n)$. In this case, the Hartree term in $E^{\rm HF}_{\downarrow}$ and $E^{\rm H}_{\uparrow}$ cancel each other and the magnon energies simplifies to 
\begin{equation}
   \mathcal{E}^m_{n} =  \int \frac{d^2 \mathbf{q}}{(2 \pi)^2} V_n(\mathbf{q}) - U_n^m
\end{equation}
The lowest spin-flip excitation energy which corresponds to the largest binding energy of the magnon ($ m = 0$) is
 \begin{equation}
      \mathcal{E}^0_{n} =  \int_{0}^{\infty}  \frac{q dq}{2 \pi} V(q) (e^{-q^2 l^2_B/2} - e^{-q^2x l^2_B})[L_n(q^2 l^2_B/2)]^2 
 \end{equation}
When $V(\mathbf{q}) > 0$, it can readily seen that $ \mathcal{E}^0_{n} >0$ in any Landau level, indicating the magnon spectrum above the fully polarized state is always gapped and hence implying robust ferromagnetism for any repulsive interaction that is spin independent. 

Next, we analyze what happens when interactions are spin-dependent ($V_n \neq U_n $) and takes the form shown in equation \eqref{eq:oppositespin_int}. As we shall show below the lowest spin-flip excitation energy above the fully polarized state is no longer guaranteed to be positive and ferromagnetism can be destabilized. 

For $d \neq 0$, the Hartree term in $E^{\rm HF}_{\downarrow}$ and $E^{\rm H}_{\uparrow}$ generally no longer cancel other and the lowest magnon energies $\mathcal{E}_n$ can become negative signaling an instability against ferromagnetism. For concreteness, let's consider again Coulomb interactions $V(\mathbf{q}) = \frac{2\pi e^2}{\epsilon |\mathbf{q}|}$, the lowest magnon energies for the $n= 0$ and $n= 1$ LL for generic values of $d$ are found to be
\begin{widetext}
\begin{equation}
\begin{aligned}
\label{eq:magnon01}
        &\mathcal{E}^0_0 = \dfrac{e^2}{\epsilon l_B} \dfrac{\sqrt{\pi}}{2}(\sqrt{2} - e^{d^2/4 l^2_B} {\rm erfc}(d/2 l_B)) + \delta E_0 \\
   & \mathcal{E}^0_1 = \dfrac{e^2}{\epsilon l_B}\frac{1}{128} \big[-12 \frac{d}{l_B} + 2 \frac{d^3}{l_B^3} + 48 \sqrt{2\pi} - (44 - 4 \frac{d^2}{l_B^2} + \frac{d^4}{l_B^2}) e^{d^2/4 l_B^2}
    \sqrt{\pi} {\rm erfc}(d/2 l_B)\big] + \delta E_1  \\
    & \delta E_n = \frac{1}{S_M} \sum_{\mathbf{k} \in {\rm BZ}  ,
 \mathbf{g}} [U_n(\mathbf{g}) - V_n(\mathbf{g})] e^{i l^2_B \mathbf{k} \wedge \mathbf{g}} 
\end{aligned}
\end{equation}
\end{widetext}

Where $\delta E_n$ is the difference between the Hartree energies of opposite-spin and equal spin interactions in the $n$-th Landau level and $\{\mathbf{g}\}$ denote reciprocal lattice vectors. We find the largest contribution to $\delta E_n$ to come from the $\mathbf{g} = 0$ part such that $\delta E_n$ can be approximated as $\delta E_n \approx \frac{-e^2 d}{\epsilon l^2_B} $. 
In Fig. \ref{fig:magnon}, we plot $\mathcal{E}_0^0$ and $\mathcal{E}_1^0$ as a function of $d/a_M$. We observe the lowest magnon energy to decrease monotonically for increasing values of $d$ until it vanishes, implying that the ferromagnetic state is no longer the true ground state. A key observation is that the lowest magnon  energy $\mathcal{E}_1^0$ for the $n = 1$ LL vanishes faster than $\mathcal{E}_0^0$ for $n = 0$ LL, meaning that ferromagnetism in 
the $n = 1$ LL is in principle less stable against the reduction of opposite-spin interaction at short distances.

We have therefore shown that spin-dependent interactions can induce a magnon instability such that the fully polarized state is no longer the ground state. Remarkably for the $n = 1$ conjugate LLs, we found that such an instability only requires a small reduction of opposite-spin interaction at very short distance ($d/a_M \approx 0.15$) than the moir\'e period, while a similar instability for the $n = 0$ conjugate LLs requires a larger spin interaction anisotropy over a larger length scale.  

\textcolor{black}{The robustness of ferromagnetism in the $n = 0$ LL relative to the $n = 1$ LL is ultimately related to the stronger repulsion at short distances for the $n = 0 $ LL, $U_0(0) > U_1 (0)$ (equation \eqref{eq:V0intra}) implying larger reduction of the opposite-spin interaction is needed in order to overcome the ferromagnetism. Following this logic, we expect ferromagnetism to be become less and less robust for higher Landau levels.  }

The magnon instability indicates a transition away from the fully polarized state towards a new ground state with different spin polarization. To figure out the true ground state of the system, we analyze next the full Hamiltonian \eqref{eq:Ham_int} across different spin sectors.

\section{VI. Phase Diagram}

\begin{figure}[t!]
    \centering
    \includegraphics[width = 0.6 \linewidth]{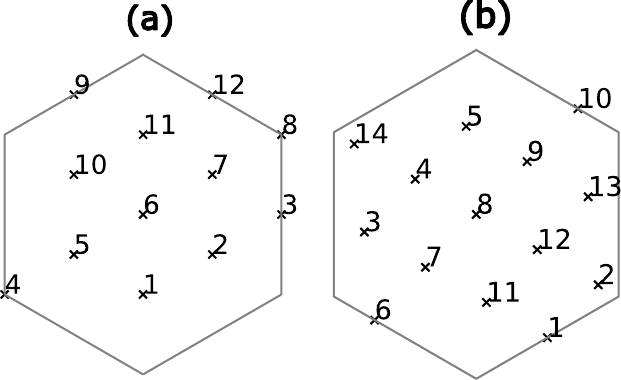}
    \caption{Finite size momentum grids used in this work. (a) 12 site cluster. (b) 14 site cluster.}
    \label{fig:clusters}
\end{figure}

We study the Hamiltonian \eqref{eq:Ham_int} with exact diagonalization (ED) on finite systems. Due to $S_z$ spin conservation, the Hamiltonian is divided into different sectors labelled by total spin $S_z = (N_{\up}-N_{\downarrow})/2$. We focus on two conjugate $n= 1$ LLs at effective band filling factor $ \nu_{\uparrow} + \nu_{\downarrow} = 1$ and consider Coulomb interactions $V(\mathbf{q}) = \frac{2\pi e^2}{\epsilon |\mathbf{q}|} $. The ED is performed on two finite momentum grids shown in Fig \ref{fig:clusters}.

To deal with the $\mathbf{q} = 0$ term in both $V_n(\mathbf{q})$ and $U_n(\mathbf{q})$, we assume the existence of an identical neutralizing charge background for both spins, as physically required. However, this doesn't fully cancel the $\mathbf{q} = 0$ contribution due the anisotropy induced by finite $d$ \cite{supp} and a constant term of the form $2 \pi e^2 d S_z^2/ (\epsilon S_M)$  \footnote{Due to our choice of interactions, our problem can be exactly mapped to a quantum Hall bilayer in opposite magnetic fields therefore this spin-dependent constant term naturally arises due to the charge imbalance of both layers.} is present and needs to be considered when comparing energies in different spin sectors \cite{macdonald_collapse_1990,zhu_exciton_2019,zhu_numerical_2017}.

\begin{figure}[t!]
    \centering
    \includegraphics[width =  \linewidth]{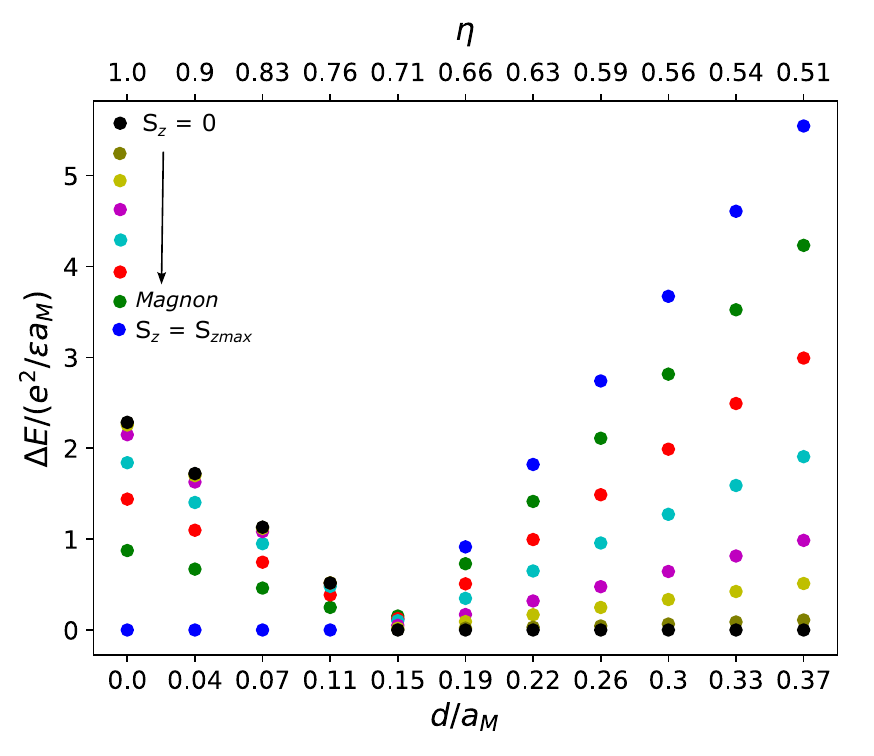}
    \caption{Phase diagram of the Hamiltonian \eqref{eq:Ham_int} for the $n = 1$ LL at effective band filling $\nu_{\uparrow} + \nu_{\downarrow} = 1$ as a function of $d/a_M$ for spin anisotropy $d$ defined in equation \eqref{eq:oppositespin_int} and $a_M$ is the moir\'e lattice constant. $S_{\rm zmax}$ is the maximum $S_z = (N_{\uparrow}-N_{\downarrow})/2$. $\eta$ is defined below equation \eqref{eq:U10}. Calculations are done on a 14 site cluser shown in Fig. \ref{fig:clusters}(b). }
    \label{fig:phasediagram}
\end{figure}

We present the phase diagram in Fig. \ref{fig:phasediagram} showing the lowest energy state in each spin sector. For values of $d/a_M < 0.15$, the lowest energy state across all sectors is the fully polarized state with $S_z = S_{\rm zmax}$, separated from other spin sectors by a finite energy gap. 

When $d$ increases, we observe the magnon gap as well as the gap to other spin sectors to decrease until it vanishes. The closure of the magnon gap happens around $d/a_M \approx 0.15$ in agreement with our exact lowest magnon energy calculation (c.f Fig. \ref{fig:magnon}). At this critical $d$ the system undergoes a \textit{direct} transition to the unpolarized state with $S_z = 0$, which remains the ground state for increasing values of $d$. It is worth noting that as the ratio of equal- to opposite-spin interactions varies, only the fully polarized and the unpolarized phases are found in our model, while partially polarized states such as considered in Refs. \cite{stefanidis_excitonic_2020, kwan_exciton_2021, kwan_excitonic_2022,villadiego2024halperin} are not ground states of our model.      

As described above, a finite $d$ parameterizes the reduction in Coulomb repulsion between opposite-spin particles 
at short distance up to $r \sim d$.  The amount of reduction can be seen from the ratio of the largest repulsion energy $U_1(0)$ (equation \eqref{eq:U10} for the $n = 1 $ LL) at $d \neq 0$ to its Coulomb value at $d = 0$ which we denote as $\eta$.  As shown in the top horizontal axis of Fig. \ref{fig:phasediagram}, the transition to the maximally unpolarized state corresponds to approximately $30 \%$ reduction of $U_1(0)$. It is remarkable that a small modification ($30\%$) of the largest repulsive Coulomb energy  corresponding to an interaction renormalization at short distance ($15\%$ of moir\'e period) is sufficient to drive the system from the fully polarized state into completely unpolarized one.

\section{VII. non-Abelian spin Hall states}

\begin{figure}[t!]
    \centering
    \includegraphics[width =1.0 \linewidth]{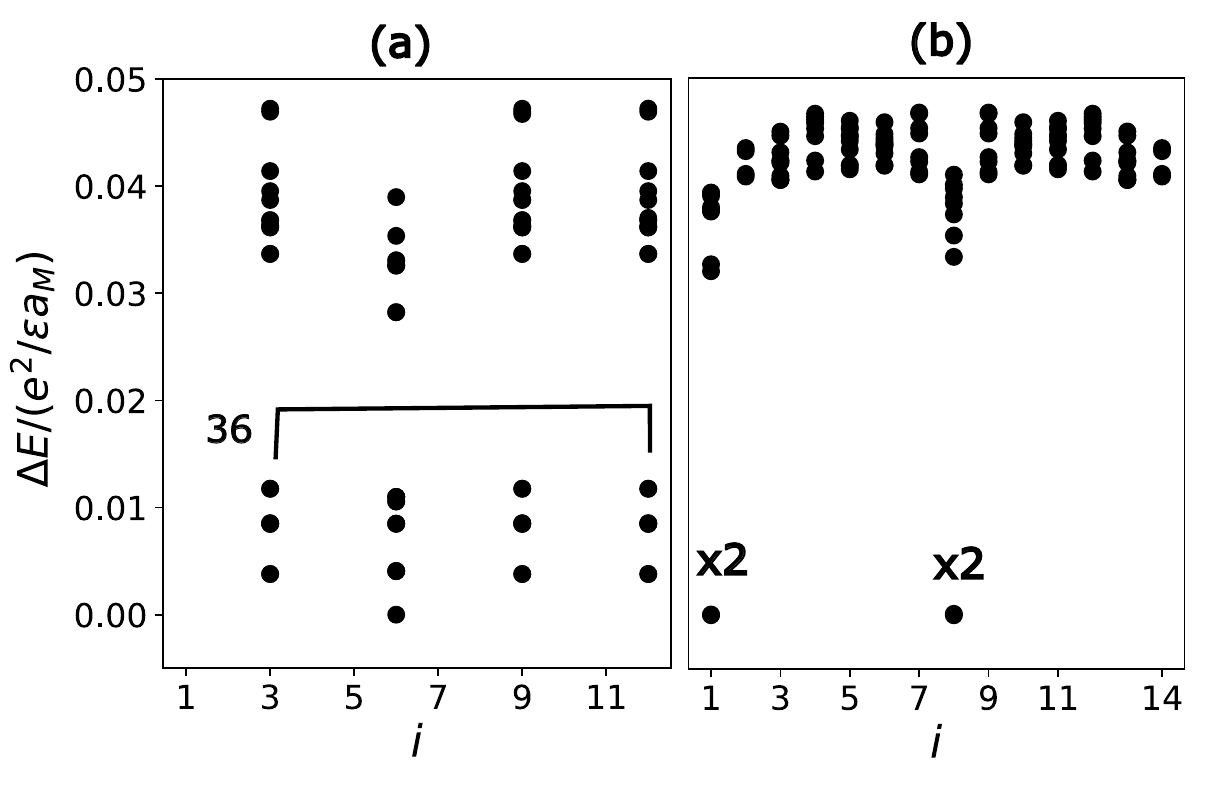}
    \caption{Many body spectrum of the Hamiltonian \eqref{eq:Ham_int}  for the $n = 1$ LL at effective band filling $\nu_{\uparrow} = \nu_{\downarrow} = 1/2$ or equivalently the $S_z = 0$ sector with $d/a_M = 0.3712 $  calculated on 12-site cluster (a) shown in Fig. \ref{fig:clusters}(a) and on 14 site cluster (b) shown in Fig. \ref{fig:clusters}(b). The index $i$ labels the momentum points of the finite cluster.}
    \label{fig:EDspectrum}
\end{figure}

After the transition to the unpolarized phase with increasing $d$, we find a set of low-lying quasi-degenerate ground states to appear.   In Fig. \ref{fig:EDspectrum}, we plot the many-body spectrum in the $S_z = 0$ sector. We observe a low-lying ground state manifold of 36 states on a 12-unit-cell system (Fig. \ref{fig:EDspectrum}(a)) and 4 states (Fig. \ref{fig:EDspectrum}(b)) on a 14-unit-cell system. This ground state degeneracy is consistent with a product of two non-abelian Moore-Read (MR) states \cite{Moore1991Aug} for spin-$\uparrow$ and $\downarrow$ particles in conjugate $n=1$ LLs, $\ket{\psi} = \ket{\psi^{\uparrow}_{\text{MR}}} \times \ket{\psi^{\downarrow}_{\text{MR}}}$, as a single MR state has a ground state degeneracy of 6 and 2 on even and odd site systems respectively \cite{ardonne2008degeneracy,Oshikawa2007Jun}.  A state $\ket{\psi}$ of this form should exhibit ground state total momenta, $K = K_{\uparrow} + K_{\downarrow}$ where $K_{\sigma}$  is the ground state total momenta for the individual $\ket{\psi^{\sigma}_{\text{MR}}}$ which is consistent with our observation.

\begin{figure}[t!]
\label{fig:gapevolution}
    \centering
    \includegraphics[width =0.8 \linewidth]{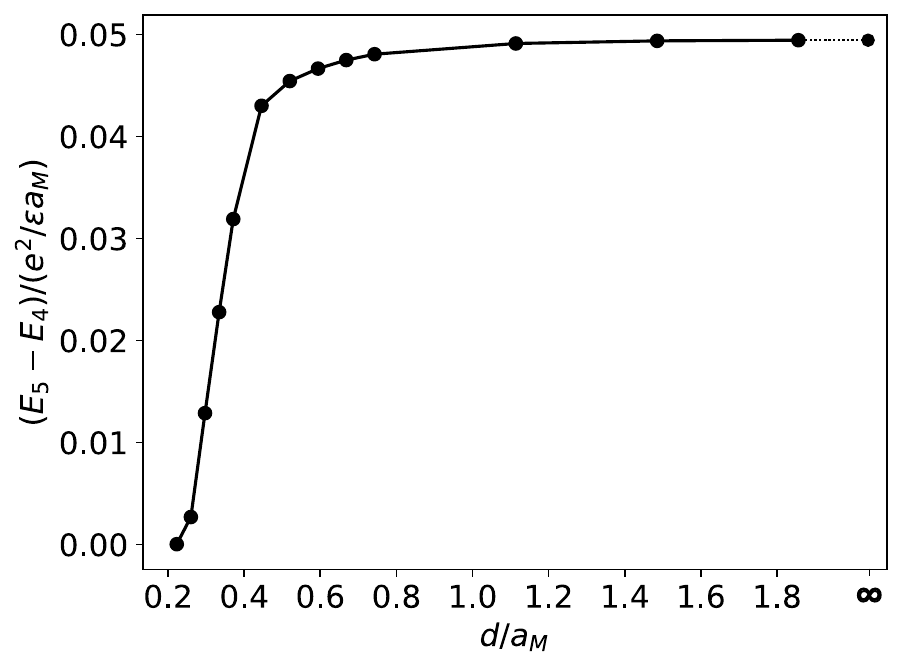}
    \caption{Energy gap $E_5 - E_4$ in the $S_z = 0$ sector of the Hamiltonian \eqref{eq:Ham_int} on the 14 site cluster (see Fig. \ref{fig:clusters}(b)) as a function of $d/a_M$. $d = \infty$ represents the decoupled spin $\uparrow$ and $\downarrow$ limit.}  
    \label{fig:energygap}
\end{figure}

In addition, we present numerical evidence that the quasi-degenerante ground states are smoothly connected to the $d \rightarrow \infty$ limit, where the opposite-spin interaction vanishes. As shown in Fig. \ref{fig:energygap}, the energy gap $E_5 - E_4$ above the ground state manifold of 14-unit-cell system remains finite. This strongly suggests that  after the spin transition, the ground state of our model stays in the same phase as two decoupled two decoupled $n = 1$ Landau levels of opposite chirality, each at half filling $\nu_{ \up} = \nu_{ \downarrow} = 1/2$.

A single $n=1$ LL with Coulomb interaction is believed to be in the non-Abelian Moore-Read phase, with the Pfaffian  and anti-Pfaffian ground states being degenerate in the thermodynamic limit due to the particle-hole symmetry, which is an exact symmetry in the absence of LL mixing \cite{lee2007particle,levin2007particle}.  In our system at  $d = \infty$, the state $\ket{\psi^{\downarrow}_{\text{MR}}}$ is related to $\ket{\psi^{\uparrow}_{\text{MR}}}$ by time-reversal symmetry: $\ket{\psi^{\downarrow}_{\text{MR}}} = \mathcal{T} \ket{\psi^{\uparrow}_{\text{MR}}}$. Therefore, we conclude from the mapping to $n=1$ LLs that our model in the decoupled limit $d=\infty$ exhibits $2\times 2=4$ types of ground states in the thermodynamic limit: 
\begin{eqnarray}
\label{eq:possiblestates}
\ket{\rm \psi_p}&=& \ket{\rm Pf}_\uparrow \times \overline{\ket{\rm Pf}}_\downarrow,  \nonumber \\
\ket{\rm \psi_a} &=& \ket{\rm aPf}_\uparrow \times \overline{\ket{\rm aPf}}_\downarrow, \nonumber \\ 
\ket{\rm \psi_+} &=& \ket{\rm Pf}_\uparrow \times \overline{\ket{\rm aPf}}_\downarrow, \nonumber \\
\ket{\rm \psi_-} &=& \ket{\rm aPf}_\uparrow \times \overline{\ket{\rm Pf}}_\downarrow, 
\end{eqnarray}
where the bar denotes time-reversal transformation. 

The above four types of states all have non-Abelian Ising anyons with charge $e/4$ in both spin-$\uparrow$ and $\downarrow$ sectors, as well as charge $e/2$ Abelian anyons \cite{Moore1991Aug,Read1999Mar,greiter1991paired,wenTopologicalOrderEdge1993}. Due to the opposite chirality of conjugate LLs, they 
have zero electrical Hall conductivity, but exhibit quantized spin Hall effect $\sigma^{ sH} = \frac{1}{2}$ and support counter-propagating charge edge modes with opposite spin. Therefore, we call these states ``non-Abelian spin Hall insulators''. 

Among these, $\ket{\rm \psi_p}$ and $\ket{\rm \psi_a}$ are time-reversal-invariant and map onto each other under particle-hole transformation. They represent ``non-Abelian topological insulator'' states at odd-integer filling, which is a non-Abelian counterpart of ``fractional topological insulators'' that are a product of a spin-$\uparrow$ Laughlin state and its time-reversal conjugate at fractional filling $\nu=2/(2m+1)$ \cite{neupert_fractional_2011, repellin_mathbbz_2_2014,wuTimereversalInvariantTopological2024}.  

In contrast, $\ket{\rm \psi_\pm}$ spontaneously break time reversal symmetry and form degenerate partners. Interestingly, despite having zero electrical Hall conductivity, this time-reversal-breaking phase exhibits a quantized thermal Hall conductance and support chiral neutral edge modes. In the presence of disorder, both of these states have 4 neutral Majorana modes propagating in the same direction giving rise to a thermal Hall conductance $\kappa_H = \pm 2$ \cite{kaneQuantizedThermalTransport1997a,Read2000Apr,cappelliThermalTransportChiral2002}. Due to its time-reversal-breaking nature, this phase may also exhibit spontaneous magnetic circular dichroism.

In the decoupled limit $d=\infty$, these four states $\ket{\rm \psi_{p,a}}$ and $\ket{\rm \psi_\pm}$ are degenerate because each half-filled LL has its own particle-hole symmetry. At finite $d$, however, the two conjugate LLs are coupled through density-density interaction, hence our band-projected Hamiltonian only has a global particle-hole symmetry. Then, we expect the degeneracy between the time-reversal-invariant states $\ket{\rm \psi_{p,a}}$ and the time-reversal breaking states $\ket{\rm \psi_{\pm}}$ becomes lifted. Thus, the true ground state of our model at finite $d>d_c$ is either a non-Abelian topological insulator, or a time-reversal breaking non-Abelian state with quantized thermal Hall effect. In either case, our system is a non-Abelian helical liquid with zero electrical Hall effect and quantized spin Hall effect.

 \section{VIII. Band mixing}

As evident from our results, the key to overcome ferromagnetism lies in the short distance part of repulsive interaction between two particles of opposite spin. A non-zero value of $d$ effectively reduces this repulsion at short distances, while it largely doesn't affect the long distance behavior ($\mathbf{q} \rightarrow 0$). We now show that this reduction arises naturally from the band mixing effect. 

In general, band mixing is expected to renormalize the underlying interactions and therefore affect the competition between different phases as shown in the study of fractional quantum Hall systems \cite{sodemann2013landau,simon_landau_2013,rezayiLandauLevelMixing2017}. In our system, since spin-$\uparrow$ and $\downarrow$ particles experience opposite magnetic fields, we may expect that interactions between two particles of equal spin and of opposite spin will be renormalized by band mixing differently. Indeed, as we show explicitly below, band mixing 
can  significantly reduce the short distance repulsion between opposite spins, which then drive the system away from ferromagnetism. 

To illustrate our point, we consider a minimum model with two time-reversed pairs of flat Chern bands: (0LL, 0$\overline{\rm LL}$) and (1LL, 1$\overline{\rm LL}$). We assume that these two spinful bands are separated by an energy gap $\Delta$, while the gap to higher bands is much larger than $\Delta$. Under such condition, it is justified to neglect the contribution from higher bands and study the band mixing effect in this two-band model for interactions smaller than $\Delta$.    
At particle filling $\nu=3$, our two-band model can be equivalently viewed as a system of holes at filling factor $\tilde{\nu}=1$, 
with the vaccuum state of holes ($\tilde{\nu}=0$) corresponding to the two spinful bands completely filled with particles ($\nu=4$).  
This hole picture is convenient for analyzing the band mixing effect below.

When the Coulomb energy scale $E_c \equiv e^2/\epsilon a_M$ is small compared to the band gap $\Delta$, 
the holes mainly reside in 1LL and 1$\overline{\rm LL}$, with a small probability of being excited to  
0LL and 0$\overline{\rm LL}$.  
To the leading order, the interaction between two holes    
in $n=1$ LL pair is simply given by the band-projected Coulomb interaction, as described earlier. The effect of band mixing 
appears at second order in  the band mixing parameter $\kappa \equiv E_c/\Delta$. By second-order perturbation theory, 
we find that 
the second order correction to the equal-spin interaction psuedopotentials due to mixing with $n = 0$ LL pair is given by 
\begin{eqnarray}
   \delta V_{1}^m = - \kappa^2 \frac{4 \pi}{\sqrt{3}}\Bigg[\dfrac{\Gamma(m+1/2) \Gamma(3/2)}{2\sqrt{\pi m! (m+1)!}}\Bigg]^2
\end{eqnarray}
while the second order correction to the opposite-spin interactions reads
\begin{widetext}
\begin{eqnarray}
        \delta U_1(\mathbf{R}^{\rm rel}) &=&  - \frac{1}{\Delta^2} \Bigg[\int \frac{d^2\mathbf{q}}{(2\pi)^2} V({\mathbf{q}}) [G^{\sigma}_{1 1}(\mathbf{q}) G^{-\sigma}_{0 1}(-\mathbf{q}) e^{i \mathbf{q} \cdot \mathbf{R}^{\rm rel}}\Bigg]^2  - \frac{1}{\Delta^2} \Bigg[ \int \frac{d^2\mathbf{q}}{(2\pi)^2} V({\mathbf{q}}) [G^{-\sigma}_{1 1}(\mathbf{q}) G^{\sigma}_{0 1}(-\mathbf{q}) ]e^{i \mathbf{q} \cdot \mathbf{R}^{\rm rel}} \Bigg]^2  \nonumber \\
        &-& \frac{1}{2 \Delta^2}\Bigg[\int \frac{d^2\mathbf{q}}{(2\pi)^2} V({\mathbf{q}})G^{\sigma}_{0 1}(\mathbf{q}) G^{-\sigma}_{0 1}(-\mathbf{q})  e^{i \mathbf{q} \cdot \mathbf{R}^{\rm rel}}\Bigg]^2
\end{eqnarray}
\end{widetext}
where $G_{n m}^{\sigma}(\mathbf{q}) = \langle n|e^{i\mathbf{q}\cdot \bar{\mathbf{R}}^{\sigma}}|m\rangle$ is form factor overlap between the $n$-th LL and $m$-th LL with $\bar{\mathbf{R}}^{\sigma}$ the cyclotron orbit coordinate for spin $\sigma$ \cite{supp}. The corrections to the opposite-spin interactions can be understood as consisting of three second-order processes: two which result from a hole with a certain spin scattering to the $n = 0$ LL while the other hole is fixed and a third process which involve two holes both scattering to the $n = 0$ LL, before scattering back to the $ n = 1$ LL. 

To put everything together, the full Coulomb two-body energies in the $n = 1 $ LL  to second order in $\kappa$ for equal-spin interactions are
    \begin{equation}
\label{eq:samespin_perturbative}
\begin{gathered}
    \tilde{V}_{1}^{m} = \dfrac{1}{\Delta} \frac{1}{\sqrt{\sqrt{3}/4\pi}} V_1^m   - \kappa^2 \frac{4 \pi}{\sqrt{3}}\Bigg[\dfrac{\Gamma(m+1/2) \Gamma(3/2)}{2\sqrt{\pi m! (m+1)!}}\Bigg]^2
    \end{gathered}
\end{equation}

Where $V_1^m$ is given in equation \eqref{eq:Haldane_1LL}. For opposite spin interactions, the expression simplifies nicely for $\mathbf{R}^{\rm rel} = 0$ and we obtain
\begin{equation}
\label{eq:oppositespin_perturbative}
\begin{split}
       \tilde{U}_{1}(0) &= \frac{1}{\Delta} \frac{1}{\sqrt{\sqrt{3}/4\pi}} U_1(0) + \delta U_1(0) \\ &=  \kappa \frac{3 \pi}{4} \sqrt{\frac{1}{2 \sqrt{3}}} - \kappa^2 \frac{4 \pi^2}{16\sqrt{3}} 
\end{split}
\end{equation}

 \begin{figure}[t!]
\label{fig:gapevolution}
    \centering
    \includegraphics[width =  \linewidth]{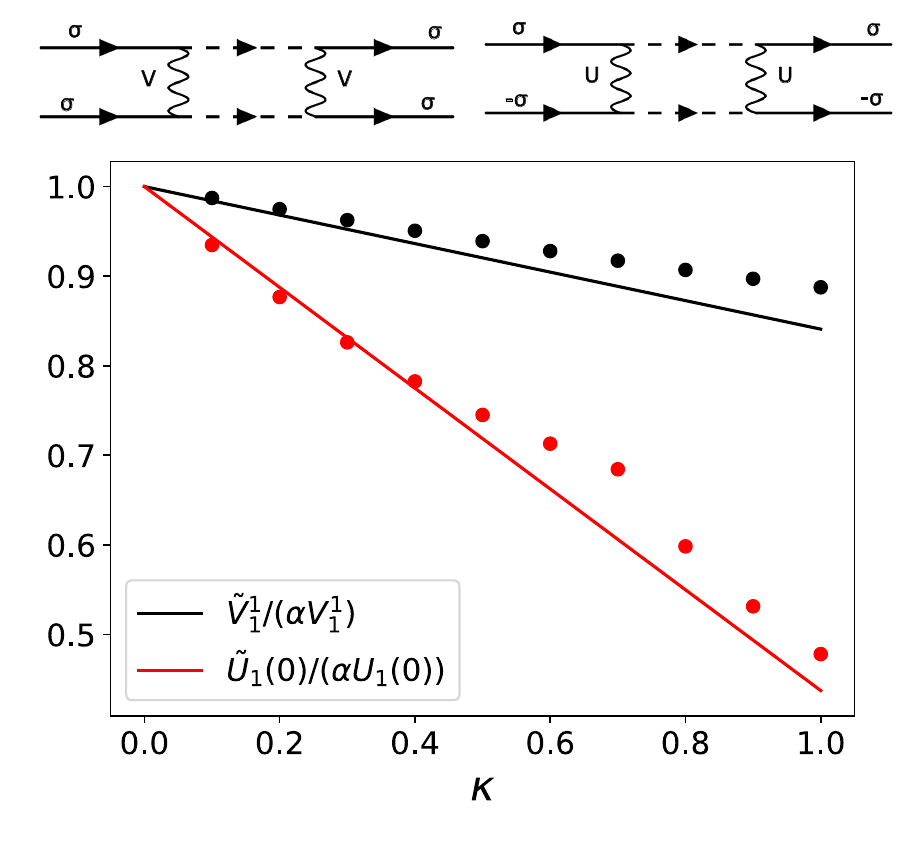}
    \caption{The largest Coulomb two body energies in the $ n = 1 $ LL calculated to second order in $\kappa$ relative to their value at first order in $\kappa$ for equal spin and opposite spin interactions as given in equations \eqref{eq:samespin_perturbative} and \eqref{eq:oppositespin_perturbative}. $\alpha = \frac{1}{\Delta} \frac{1}{\sqrt{\sqrt{3}/4\pi}} $. The dots are the numerically obtained two-body energies from exact diagonalization of the two-band model discussed \textcolor{black}{on a $6 \times 6$ lattice} \cite{supp}. The top panel represents a schematic of a second order process that renormalizes equal-spin and opposite-spin  interactions. Two holes with equal-spin (left) or opposite spin (right) scatter from the $ n = 1$ LL (the solid lines) to the $ n = 0$ LL (the dashed lines) then scatter back to the $ n = 1$ LL. } 
    \label{fig:bandmixing}
\end{figure}

With $U_1(0)$ given in equation \eqref{eq:U10} with $d = 0$. In Fig. \ref{fig:bandmixing}., we plot the  the largest repulsive components for both equal-spin ($ m = 1$) and opposite-spin interactions at second order in $\kappa$ relative to their values at first order in $\kappa$. We observe that opposite-spin interactions gets strongly suppressed due to band mixing. Remarkably, the reduction of opposite-spin interactions within this two-band model already reaches the needed value to drive the system out of ferromagnetism ($\eta \approx 0.71$ in Fig. \ref{fig:phasediagram}). While same-spin interactions gets also suppressed by band mixing, this suppression is clearly smaller than the opposite-spin case. 

To confirm the validity of our perturbative calculations, we numerically diagonalize the full two-hole problem in our two-band model \cite{supp}. As shown in Fig. \ref{fig:bandmixing}, the numerically obtained two-body spectrum (the plotted dots) also shows strong suppression of the opposite-spin interactions due to band mixing in addition to a much weaker suppression for  equal-spin interactions. For smaller values of $\kappa$, the numerical  two-body energies displays excellent agreement with the values obtained perturbatively.

We have therefore established band mixing as a generic mechanism to renormalize interactions at short-distances which effectively results in a significant reduction of the opposite-spin interactions compared to equal-spin interactions even when the bare values of both are the same.

\section{IX. Twisted TMD bilayers at $\nu=3$}

We now discuss the relevance of our theory for twisted semiconductor bilayers. Both $t$MoTe$_2$ and $t$WSe$_2$ possess conjugate Chern bands with opposite spin \cite{wu_topological_2019,devakul2021magic}. In a range of small twist angles, transport experiments have observed the quantum spin Hall effect at $\nu=2$ and the double quantum spin Hall effect at $\nu=4$ \cite{kang_evidence_2024,kang2024observationWSe2}, consistent with the theoretical prediction that the lowest two moir\'e bands have the same spin Chern number \cite{devakul2021magic,reddy2023fractional,zhangPolarizationdrivenBandTopology2024}. Moreover, near the magic angle, the approximate mapping from the first or second moir\'e bands to the $n=0,1$ LL has been theoretically suggested \cite{morales2023magic, reddy2024non,xu2024multiple,wang2024higher,ahn2024first,chen2024robust}. Therefore, despite its simplicity, our two-band model correctly captures the band topology for $t$TMD as well as the most essential feature of its moir\'e bands. 

As evident from our magnon gap calculations (Fig. \ref{fig:magnon}), ferromagnetism is more robust in the $n = 0$ LL and survives larger values of spin interactions anisotropy. This can be intuitively understood due to the fact that short range repulsion, quantified by the Haldane pseudopotentials, is stronger in the $n = 0$ LL than the $n = 1$ LL. The absence of the quantum anomalous Hall effect at $\nu = 3$ \cite{kang_evidence_2024} in contrast to $\nu = 1 $ is therefore consistent with this observation. We also note that band mixing effects have been shown \cite{abouelkomsan2024band,wubandmixing2024} to be  significant in twisted TMDs. As we have shown, band mixing provides a possible mechanism for opposite-spin reduction which can drive the system away from ferromagentism.

The non-Abelian spin Hall insulators we found in our model have an insulating gap in the bulk and zero electrical Hall conductivity. Due to the doubled Moore-Read topological order, they have counter-propagating charge edge mode that are topologically protected in the presence of spin-$S_z$ conservation. These features are consistent with the recent transport experiment on a $t$MoTe$_2$ device at $\theta=2.1^\circ$, which found vanishing electrical Hall conductivity at $\nu=3$, accompanied with fractional edge conductance of $3 e^2/(2h)$ \textcolor{black}{per spin} over micron length scale \cite{kang_evidence_2024}.

Through a microscopic model study, we have identified two types of non-Abelian spin Hall insulators, which are time-reversal-invariant and time-reversal breaking respectively. Generally speaking, spontaneous time-reversal breaking can be detected by magnetic circular dichroism, which measures the imaginary part of Hall conductivity at optical frequency. In addition, our time-reversal-breaking non-Abelian state exhibits quantized thermal Hall conductance and supports chiral neutral edge modes. 

It is worth noting that the topological order of our non-Abelian spin Hall insulators does not rely on any symmetry. Irrespective of whether time-reversal or spin-$S_z$ symmetry is present or not, as long as charge conservation is maintained, 
these non-Abelian insulators have charge $e/4$ non-Abelian anyons and, whereas recently proposed Abelian states have  minimally charged anyons with charge $e/2$ \cite{jian2024minimal, zhang2024non, villadiego2024halperin}. Therefore, at least in principle, the presence or absence of non-Abelian fractionalization in $t$MoTe$_2$ can be be distinguished by measuring the fractional charge in tunneling, charge sensing and interferometer experiments.

\section{X. Discussion}

Our model study leaves a number of important questions open for further study. The states $\ket{\psi_{\pm}}$  we identified represent a novel class of time-reversal breaking non-Abelian topological order with zero electrical Hall conductivity.  Whether these states or the time-reversal-invariant  ($\ket{\psi_{p,a}}$)  are the true ground states can be addressed at large $d$ by treating opposite-spin interaction as a perturbation to two decoupled conjugate $n=1$ LLs. In addition, the degeneracy between time-reversal-invariant Pfaffian and anti-Pfaffian states, $\ket{\psi_{p}}$ and $\ket{\psi_{a}}$, should be lifted by band mixing, which generates three-body interactions that break particle-hole symmetry.

Regarding the time-reversal-invariant non-Abelian spin Hall insulator, an important issue is the stability of its edge state against $S_z$-breaking and time-reversal-invariant perturbations \cite{may2024theory,cappelliStabilityTopologicalInsulators2015} (such as Rashba spin-orbit interaction), which are expected to appear in twisted semiconductors at the edge or due to disorder. If the gapless edge states are stable in the presence of time-reversal symmetry, the non-Abelian spin Hall insulator can be regarded as a non-Abelian fractional topological insulator.              

Despite its simplicity, our model is expected to host a plethora of topological phases at various fillings and under magnetic field. At high field, our model at $\nu=3$ will be a Chern insulator because spin-polarized particles completely fill the second Chern band. This is consistent with the recent transport experiment on $t$MoTe$_2$ at $2.6^\circ$ \cite{park2024ferromagnetism}. A detailed numerical study of our model will be reported in future work. 

In our analysis, we have solely considered the flat band limit. While the fully polarized state is completely insensitive to effects from non-interacting or interaction-induced dispersion, the energies of partially polarized phases can be lowered by dispersion which implies a larger window of non-ferromagnetic phases. However, dispersion can disfavor strongly interacting phases such as the non-Abelian spin Hall insulator. 

Because of the underyling LL wavefunctions, our model exhibits uniform band geometry. Another important question concerns the possible effects of non-uniform band geometry which are present in twisted semiconductor bilayers. For instance, a interesting future direction is investigating the stability of ferromagnetism against band geometry fluctuations which modify the band projected interactions.

\textcolor{black}{We note that the non-Abelian spin Hall phase found within our model of two conjugate LLs is uniquely identified by its topological ground state degeneracy---36 for even number of particles and 4 for odd, and therefore is different from recently proposed phases \cite{jian2024minimal, zhang2024non} which have different ground state degeneracies. In addition, our state has zero total spin $S_z=0$ and thus is distinct from those considered in Ref.\cite{villadiego2024halperin} which have partial spin polarization.} 

Finally, while we have focused on total filling $\nu_{\uparrow} + \nu_{\downarrow} = 1$, our analysis naturally extends to other filling fractions that could support partially or completely unpolarized topologically ordered phases. In particular, we have identified band mixing as a universal microscopic mechanism for opposite-spin reduction which is a necessary ingredient for the stabilization of such phases.

\section{Acknowledgements}
\begin{acknowledgements}
We thank Aidan Reddy and Nisarga Paul for a related collaboration on non-Abelian fractional Chern insulators \cite{reddy2024non}. 
It is a pleasure to thank Kin Fai Mak, Jie Shan and Xiaodong Xu for insightful discussions on experiments.  
This work is supported by  the U.S. Army Research Laboratory and the U.S. Army Research Office through the
 Institute for Soldier Nanotechnologies, under Collaborative Agreement No. W911NF-18-2-0048. A.A. was supported by the Knut and
Alice Wallenberg Foundation (KAW 2022.0348).  L.F. was partly supported by the Simons Investigator Award from the Simons Foundation. The authors acknowledge the MIT SuperCloud and Lincoln Laboratory Supercomputing Center for providing computing resources. 
\end{acknowledgements}

\bibliography{refs}

\begin{thebibliography}{74}%
\makeatletter
\providecommand \@ifxundefined [1]{%
 \@ifx{#1\undefined}
}%
\providecommand \@ifnum [1]{%
 \ifnum #1\expandafter \@firstoftwo
 \else \expandafter \@secondoftwo
 \fi
}%
\providecommand \@ifx [1]{%
 \ifx #1\expandafter \@firstoftwo
 \else \expandafter \@secondoftwo
 \fi
}%
\providecommand \natexlab [1]{#1}%
\providecommand \enquote  [1]{``#1''}%
\providecommand \bibnamefont  [1]{#1}%
\providecommand \bibfnamefont [1]{#1}%
\providecommand \citenamefont [1]{#1}%
\providecommand \href@noop [0]{\@secondoftwo}%
\providecommand \href [0]{\begingroup \@sanitize@url \@href}%
\providecommand \@href[1]{\@@startlink{#1}\@@href}%
\providecommand \@@href[1]{\endgroup#1\@@endlink}%
\providecommand \@sanitize@url [0]{\catcode `\\12\catcode `\$12\catcode `\&12\catcode `\#12\catcode `\^12\catcode `\_12\catcode `\%12\relax}%
\providecommand \@@startlink[1]{}%
\providecommand \@@endlink[0]{}%
\providecommand \url  [0]{\begingroup\@sanitize@url \@url }%
\providecommand \@url [1]{\endgroup\@href {#1}{\urlprefix }}%
\providecommand \urlprefix  [0]{URL }%
\providecommand \Eprint [0]{\href }%
\providecommand \doibase [0]{http://dx.doi.org/}%
\providecommand \selectlanguage [0]{\@gobble}%
\providecommand \bibinfo  [0]{\@secondoftwo}%
\providecommand \bibfield  [0]{\@secondoftwo}%
\providecommand \translation [1]{[#1]}%
\providecommand \BibitemOpen [0]{}%
\providecommand \bibitemStop [0]{}%
\providecommand \bibitemNoStop [0]{.\EOS\space}%
\providecommand \EOS [0]{\spacefactor3000\relax}%
\providecommand \BibitemShut  [1]{\csname bibitem#1\endcsname}%
\let\auto@bib@innerbib\@empty
\bibitem [{\citenamefont {Tsui}\ \emph {et~al.}(1982)\citenamefont {Tsui}, \citenamefont {Stormer},\ and\ \citenamefont {Gossard}}]{tsui1982two}%
  \BibitemOpen
  \bibfield  {author} {\bibinfo {author} {\bibfnamefont {D.~C.}\ \bibnamefont {Tsui}}, \bibinfo {author} {\bibfnamefont {H.~L.}\ \bibnamefont {Stormer}}, \ and\ \bibinfo {author} {\bibfnamefont {A.~C.}\ \bibnamefont {Gossard}},\ }\bibfield  {title} {\enquote {\bibinfo {title} {Two-dimensional magnetotransport in the extreme quantum limit},}\ }\href {\doibase 10.1103/PhysRevLett.48.1559} {\bibfield  {journal} {\bibinfo  {journal} {Phys. Rev. Lett.}\ }\textbf {\bibinfo {volume} {48}},\ \bibinfo {pages} {1559--1562} (\bibinfo {year} {1982})}\BibitemShut {NoStop}%
\bibitem [{\citenamefont {Laughlin}(1983)}]{laughlin_anomalous_1983}%
  \BibitemOpen
  \bibfield  {author} {\bibinfo {author} {\bibfnamefont {R.~B.}\ \bibnamefont {Laughlin}},\ }\bibfield  {title} {\enquote {\bibinfo {title} {Anomalous {Quantum} {Hall} {Effect}: {An} {Incompressible} {Quantum} {Fluid} with {Fractionally} {Charged} {Excitations}},}\ }\href {\doibase 10.1103/PhysRevLett.50.1395} {\bibfield  {journal} {\bibinfo  {journal} {Physical Review Letters}\ }\textbf {\bibinfo {volume} {50}},\ \bibinfo {pages} {1395--1398} (\bibinfo {year} {1983})}\BibitemShut {NoStop}%
\bibitem [{\citenamefont {Zeng}\ \emph {et~al.}(2023)\citenamefont {Zeng}, \citenamefont {Xia}, \citenamefont {Kang}, \citenamefont {Zhu}, \citenamefont {Kn{\"u}ppel}, \citenamefont {Vaswani}, \citenamefont {Watanabe}, \citenamefont {Taniguchi}, \citenamefont {Mak},\ and\ \citenamefont {Shan}}]{zeng2023thermodynamic}%
  \BibitemOpen
  \bibfield  {author} {\bibinfo {author} {\bibfnamefont {Yihang}\ \bibnamefont {Zeng}}, \bibinfo {author} {\bibfnamefont {Zhengchao}\ \bibnamefont {Xia}}, \bibinfo {author} {\bibfnamefont {Kaifei}\ \bibnamefont {Kang}}, \bibinfo {author} {\bibfnamefont {Jiacheng}\ \bibnamefont {Zhu}}, \bibinfo {author} {\bibfnamefont {Patrick}\ \bibnamefont {Kn{\"u}ppel}}, \bibinfo {author} {\bibfnamefont {Chirag}\ \bibnamefont {Vaswani}}, \bibinfo {author} {\bibfnamefont {Kenji}\ \bibnamefont {Watanabe}}, \bibinfo {author} {\bibfnamefont {Takashi}\ \bibnamefont {Taniguchi}}, \bibinfo {author} {\bibfnamefont {Kin~Fai}\ \bibnamefont {Mak}}, \ and\ \bibinfo {author} {\bibfnamefont {Jie}\ \bibnamefont {Shan}},\ }\bibfield  {title} {\enquote {\bibinfo {title} {Thermodynamic evidence of fractional chern insulator in moir{\'e} mote2},}\ }\href {https://www.nature.com/articles/s41586-023-06452-3} {\bibfield  {journal} {\bibinfo  {journal} {Nature}\ ,\ \bibinfo {pages} {1--2}} (\bibinfo {year} {2023})}\BibitemShut {NoStop}%
\bibitem [{\citenamefont {Cai}\ \emph {et~al.}(2023)\citenamefont {Cai}, \citenamefont {Anderson}, \citenamefont {Wang}, \citenamefont {Zhang}, \citenamefont {Liu}, \citenamefont {Holtzmann}, \citenamefont {Zhang}, \citenamefont {Fan}, \citenamefont {Taniguchi}, \citenamefont {Watanabe} \emph {et~al.}}]{cai2023signatures}%
  \BibitemOpen
  \bibfield  {author} {\bibinfo {author} {\bibfnamefont {Jiaqi}\ \bibnamefont {Cai}}, \bibinfo {author} {\bibfnamefont {Eric}\ \bibnamefont {Anderson}}, \bibinfo {author} {\bibfnamefont {Chong}\ \bibnamefont {Wang}}, \bibinfo {author} {\bibfnamefont {Xiaowei}\ \bibnamefont {Zhang}}, \bibinfo {author} {\bibfnamefont {Xiaoyu}\ \bibnamefont {Liu}}, \bibinfo {author} {\bibfnamefont {William}\ \bibnamefont {Holtzmann}}, \bibinfo {author} {\bibfnamefont {Yinong}\ \bibnamefont {Zhang}}, \bibinfo {author} {\bibfnamefont {Fengren}\ \bibnamefont {Fan}}, \bibinfo {author} {\bibfnamefont {Takashi}\ \bibnamefont {Taniguchi}}, \bibinfo {author} {\bibfnamefont {Kenji}\ \bibnamefont {Watanabe}},  \emph {et~al.},\ }\bibfield  {title} {\enquote {\bibinfo {title} {Signatures of fractional quantum anomalous hall states in twisted mote2},}\ }\href {https://doi.org/10.1038/s41586-023-06289-w} {\bibfield  {journal} {\bibinfo  {journal} {Nature}\ ,\ \bibinfo {pages} {1--3}} (\bibinfo {year} {2023})}\BibitemShut {NoStop}%
\bibitem [{\citenamefont {Park}\ \emph {et~al.}(2023)\citenamefont {Park}, \citenamefont {Cai}, \citenamefont {Anderson}, \citenamefont {Zhang}, \citenamefont {Zhu}, \citenamefont {Liu}, \citenamefont {Wang}, \citenamefont {Holtzmann}, \citenamefont {Hu}, \citenamefont {Liu} \emph {et~al.}}]{park2023observation}%
  \BibitemOpen
  \bibfield  {author} {\bibinfo {author} {\bibfnamefont {Heonjoon}\ \bibnamefont {Park}}, \bibinfo {author} {\bibfnamefont {Jiaqi}\ \bibnamefont {Cai}}, \bibinfo {author} {\bibfnamefont {Eric}\ \bibnamefont {Anderson}}, \bibinfo {author} {\bibfnamefont {Yinong}\ \bibnamefont {Zhang}}, \bibinfo {author} {\bibfnamefont {Jiayi}\ \bibnamefont {Zhu}}, \bibinfo {author} {\bibfnamefont {Xiaoyu}\ \bibnamefont {Liu}}, \bibinfo {author} {\bibfnamefont {Chong}\ \bibnamefont {Wang}}, \bibinfo {author} {\bibfnamefont {William}\ \bibnamefont {Holtzmann}}, \bibinfo {author} {\bibfnamefont {Chaowei}\ \bibnamefont {Hu}}, \bibinfo {author} {\bibfnamefont {Zhaoyu}\ \bibnamefont {Liu}},  \emph {et~al.},\ }\bibfield  {title} {\enquote {\bibinfo {title} {Observation of fractionally quantized anomalous hall effect},}\ }\href {https://doi.org/10.1038/s41586-023-06536-0} {\bibfield  {journal} {\bibinfo  {journal} {Nature}\ ,\ \bibinfo {pages} {1--3}} (\bibinfo {year} {2023})}\BibitemShut {NoStop}%
\bibitem [{\citenamefont {Xu}\ \emph {et~al.}(2023)\citenamefont {Xu}, \citenamefont {Sun}, \citenamefont {Jia}, \citenamefont {Liu}, \citenamefont {Xu}, \citenamefont {Li}, \citenamefont {Gu}, \citenamefont {Watanabe}, \citenamefont {Taniguchi}, \citenamefont {Tong}, \citenamefont {Jia}, \citenamefont {Shi}, \citenamefont {Jiang}, \citenamefont {Zhang}, \citenamefont {Liu},\ and\ \citenamefont {Li}}]{xu_observation_2023}%
  \BibitemOpen
  \bibfield  {author} {\bibinfo {author} {\bibfnamefont {Fan}\ \bibnamefont {Xu}}, \bibinfo {author} {\bibfnamefont {Zheng}\ \bibnamefont {Sun}}, \bibinfo {author} {\bibfnamefont {Tongtong}\ \bibnamefont {Jia}}, \bibinfo {author} {\bibfnamefont {Chang}\ \bibnamefont {Liu}}, \bibinfo {author} {\bibfnamefont {Cheng}\ \bibnamefont {Xu}}, \bibinfo {author} {\bibfnamefont {Chushan}\ \bibnamefont {Li}}, \bibinfo {author} {\bibfnamefont {Yu}~\bibnamefont {Gu}}, \bibinfo {author} {\bibfnamefont {Kenji}\ \bibnamefont {Watanabe}}, \bibinfo {author} {\bibfnamefont {Takashi}\ \bibnamefont {Taniguchi}}, \bibinfo {author} {\bibfnamefont {Bingbing}\ \bibnamefont {Tong}}, \bibinfo {author} {\bibfnamefont {Jinfeng}\ \bibnamefont {Jia}}, \bibinfo {author} {\bibfnamefont {Zhiwen}\ \bibnamefont {Shi}}, \bibinfo {author} {\bibfnamefont {Shengwei}\ \bibnamefont {Jiang}}, \bibinfo {author} {\bibfnamefont {Yang}\ \bibnamefont {Zhang}}, \bibinfo {author} {\bibfnamefont {Xiaoxue}\ \bibnamefont {Liu}}, \ and\ \bibinfo {author}
  {\bibfnamefont {Tingxin}\ \bibnamefont {Li}},\ }\bibfield  {title} {\enquote {\bibinfo {title} {Observation of {Integer} and {Fractional} {Quantum} {Anomalous} {Hall} {Effects} in {Twisted} {Bilayer} {${\rm MoTe}_2$}},}\ }\href {\doibase 10.1103/PhysRevX.13.031037} {\bibfield  {journal} {\bibinfo  {journal} {Physical Review X}\ }\textbf {\bibinfo {volume} {13}},\ \bibinfo {pages} {031037} (\bibinfo {year} {2023})}\BibitemShut {NoStop}%
\bibitem [{\citenamefont {Lu}\ \emph {et~al.}(2024)\citenamefont {Lu}, \citenamefont {Han}, \citenamefont {Yao}, \citenamefont {Reddy}, \citenamefont {Yang}, \citenamefont {Seo}, \citenamefont {Watanabe}, \citenamefont {Taniguchi}, \citenamefont {Fu},\ and\ \citenamefont {Ju}}]{Lu2024Feb}%
  \BibitemOpen
  \bibfield  {author} {\bibinfo {author} {\bibfnamefont {Zhengguang}\ \bibnamefont {Lu}}, \bibinfo {author} {\bibfnamefont {Tonghang}\ \bibnamefont {Han}}, \bibinfo {author} {\bibfnamefont {Yuxuan}\ \bibnamefont {Yao}}, \bibinfo {author} {\bibfnamefont {Aidan~P.}\ \bibnamefont {Reddy}}, \bibinfo {author} {\bibfnamefont {Jixiang}\ \bibnamefont {Yang}}, \bibinfo {author} {\bibfnamefont {Junseok}\ \bibnamefont {Seo}}, \bibinfo {author} {\bibfnamefont {Kenji}\ \bibnamefont {Watanabe}}, \bibinfo {author} {\bibfnamefont {Takashi}\ \bibnamefont {Taniguchi}}, \bibinfo {author} {\bibfnamefont {Liang}\ \bibnamefont {Fu}}, \ and\ \bibinfo {author} {\bibfnamefont {Long}\ \bibnamefont {Ju}},\ }\bibfield  {title} {\enquote {\bibinfo {title} {{Fractional quantum anomalous Hall effect in multilayer graphene}},}\ }\href {\doibase 10.1038/s41586-023-07010-7} {\bibfield  {journal} {\bibinfo  {journal} {Nature}\ }\textbf {\bibinfo {volume} {626}},\ \bibinfo {pages} {759--764} (\bibinfo {year} {2024})}\BibitemShut {NoStop}%
\bibitem [{\citenamefont {Li}\ \emph {et~al.}(2021)\citenamefont {Li}, \citenamefont {Kumar}, \citenamefont {Sun},\ and\ \citenamefont {Lin}}]{li2021spontaneous}%
  \BibitemOpen
  \bibfield  {author} {\bibinfo {author} {\bibfnamefont {Heqiu}\ \bibnamefont {Li}}, \bibinfo {author} {\bibfnamefont {Umesh}\ \bibnamefont {Kumar}}, \bibinfo {author} {\bibfnamefont {Kai}\ \bibnamefont {Sun}}, \ and\ \bibinfo {author} {\bibfnamefont {Shi-Zeng}\ \bibnamefont {Lin}},\ }\bibfield  {title} {\enquote {\bibinfo {title} {Spontaneous fractional chern insulators in transition metal dichalcogenide moir{\'e} superlattices},}\ }\href {https://doi.org/10.1103/PhysRevResearch.3.L032070} {\bibfield  {journal} {\bibinfo  {journal} {Physical Review Research}\ }\textbf {\bibinfo {volume} {3}},\ \bibinfo {pages} {L032070} (\bibinfo {year} {2021})}\BibitemShut {NoStop}%
\bibitem [{\citenamefont {Cr{\'e}pel}\ and\ \citenamefont {Fu}(2023)}]{crepel2023anomalous}%
  \BibitemOpen
  \bibfield  {author} {\bibinfo {author} {\bibfnamefont {Valentin}\ \bibnamefont {Cr{\'e}pel}}\ and\ \bibinfo {author} {\bibfnamefont {Liang}\ \bibnamefont {Fu}},\ }\bibfield  {title} {\enquote {\bibinfo {title} {Anomalous hall metal and fractional chern insulator in twisted transition metal dichalcogenides},}\ }\href {https://doi.org/10.1103/PhysRevB.107.L201109} {\bibfield  {journal} {\bibinfo  {journal} {Physical Review B}\ }\textbf {\bibinfo {volume} {107}},\ \bibinfo {pages} {L201109} (\bibinfo {year} {2023})}\BibitemShut {NoStop}%
\bibitem [{\citenamefont {Levin}\ and\ \citenamefont {Stern}(2009)}]{levin_fractional_2009}%
  \BibitemOpen
  \bibfield  {author} {\bibinfo {author} {\bibfnamefont {Michael}\ \bibnamefont {Levin}}\ and\ \bibinfo {author} {\bibfnamefont {Ady}\ \bibnamefont {Stern}},\ }\bibfield  {title} {\enquote {\bibinfo {title} {Fractional {Topological} {Insulators}},}\ }\href {\doibase 10.1103/PhysRevLett.103.196803} {\bibfield  {journal} {\bibinfo  {journal} {Physical Review Letters}\ }\textbf {\bibinfo {volume} {103}},\ \bibinfo {pages} {196803} (\bibinfo {year} {2009})}\BibitemShut {NoStop}%
\bibitem [{\citenamefont {Neupert}\ \emph {et~al.}(2011)\citenamefont {Neupert}, \citenamefont {Santos}, \citenamefont {Ryu}, \citenamefont {Chamon},\ and\ \citenamefont {Mudry}}]{neupert_fractional_2011}%
  \BibitemOpen
  \bibfield  {author} {\bibinfo {author} {\bibfnamefont {Titus}\ \bibnamefont {Neupert}}, \bibinfo {author} {\bibfnamefont {Luiz}\ \bibnamefont {Santos}}, \bibinfo {author} {\bibfnamefont {Shinsei}\ \bibnamefont {Ryu}}, \bibinfo {author} {\bibfnamefont {Claudio}\ \bibnamefont {Chamon}}, \ and\ \bibinfo {author} {\bibfnamefont {Christopher}\ \bibnamefont {Mudry}},\ }\bibfield  {title} {\enquote {\bibinfo {title} {Fractional topological liquids with time-reversal symmetry and their lattice realization},}\ }\href {\doibase 10.1103/PhysRevB.84.165107} {\bibfield  {journal} {\bibinfo  {journal} {Physical Review B}\ }\textbf {\bibinfo {volume} {84}},\ \bibinfo {pages} {165107} (\bibinfo {year} {2011})}\BibitemShut {NoStop}%
\bibitem [{\citenamefont {Freedman}\ \emph {et~al.}(2004)\citenamefont {Freedman}, \citenamefont {Nayak}, \citenamefont {Shtengel}, \citenamefont {Walker},\ and\ \citenamefont {Wang}}]{freedman_class_2004}%
  \BibitemOpen
  \bibfield  {author} {\bibinfo {author} {\bibfnamefont {Michael}\ \bibnamefont {Freedman}}, \bibinfo {author} {\bibfnamefont {Chetan}\ \bibnamefont {Nayak}}, \bibinfo {author} {\bibfnamefont {Kirill}\ \bibnamefont {Shtengel}}, \bibinfo {author} {\bibfnamefont {Kevin}\ \bibnamefont {Walker}}, \ and\ \bibinfo {author} {\bibfnamefont {Zhenghan}\ \bibnamefont {Wang}},\ }\bibfield  {title} {\enquote {\bibinfo {title} {A class of \textit{{P}},\textit{{T}}-invariant topological phases of interacting electrons},}\ }\href {\doibase 10.1016/j.aop.2004.01.006} {\bibfield  {journal} {\bibinfo  {journal} {Annals of Physics}\ }\textbf {\bibinfo {volume} {310}},\ \bibinfo {pages} {428--492} (\bibinfo {year} {2004})}\BibitemShut {NoStop}%
\bibitem [{\citenamefont {Repellin}\ \emph {et~al.}(2014)\citenamefont {Repellin}, \citenamefont {Bernevig},\ and\ \citenamefont {Regnault}}]{repellin_mathbbz_2_2014}%
  \BibitemOpen
  \bibfield  {author} {\bibinfo {author} {\bibfnamefont {C.}~\bibnamefont {Repellin}}, \bibinfo {author} {\bibfnamefont {B.~Andrei}\ \bibnamefont {Bernevig}}, \ and\ \bibinfo {author} {\bibfnamefont {N.}~\bibnamefont {Regnault}},\ }\bibfield  {title} {\enquote {\bibinfo {title} {{${\rm Z_2}$} {fractional} {topological} {insulators} in {two} {dimensions}},}\ }\href {\doibase 10.1103/PhysRevB.90.245401} {\bibfield  {journal} {\bibinfo  {journal} {Physical Review B}\ }\textbf {\bibinfo {volume} {90}},\ \bibinfo {pages} {245401} (\bibinfo {year} {2014})}\BibitemShut {NoStop}%
\bibitem [{\citenamefont {Bernevig}\ and\ \citenamefont {Zhang}(2006)}]{bernevigQuantumSpinHall2006}%
  \BibitemOpen
  \bibfield  {author} {\bibinfo {author} {\bibfnamefont {B.~Andrei}\ \bibnamefont {Bernevig}}\ and\ \bibinfo {author} {\bibfnamefont {Shou-Cheng}\ \bibnamefont {Zhang}},\ }\bibfield  {title} {\enquote {\bibinfo {title} {Quantum {{Spin Hall Effect}}},}\ }\href {\doibase 10.1103/PhysRevLett.96.106802} {\bibfield  {journal} {\bibinfo  {journal} {Physical Review Letters}\ }\textbf {\bibinfo {volume} {96}},\ \bibinfo {pages} {106802} (\bibinfo {year} {2006})}\BibitemShut {NoStop}%
\bibitem [{\citenamefont {Kane}\ and\ \citenamefont {Mele}(2005)}]{kane2005quantum}%
  \BibitemOpen
  \bibfield  {author} {\bibinfo {author} {\bibfnamefont {Charles~L}\ \bibnamefont {Kane}}\ and\ \bibinfo {author} {\bibfnamefont {Eugene~J}\ \bibnamefont {Mele}},\ }\bibfield  {title} {\enquote {\bibinfo {title} {Quantum spin hall effect in graphene},}\ }\href {https://doi.org/10.1103/PhysRevLett.95.226801} {\bibfield  {journal} {\bibinfo  {journal} {Physical review letters}\ }\textbf {\bibinfo {volume} {95}},\ \bibinfo {pages} {226801} (\bibinfo {year} {2005})}\BibitemShut {NoStop}%
\bibitem [{\citenamefont {Wu}\ \emph {et~al.}(2019)\citenamefont {Wu}, \citenamefont {Lovorn}, \citenamefont {Tutuc}, \citenamefont {Martin},\ and\ \citenamefont {MacDonald}}]{wu_topological_2019}%
  \BibitemOpen
  \bibfield  {author} {\bibinfo {author} {\bibfnamefont {Fengcheng}\ \bibnamefont {Wu}}, \bibinfo {author} {\bibfnamefont {Timothy}\ \bibnamefont {Lovorn}}, \bibinfo {author} {\bibfnamefont {Emanuel}\ \bibnamefont {Tutuc}}, \bibinfo {author} {\bibfnamefont {Ivar}\ \bibnamefont {Martin}}, \ and\ \bibinfo {author} {\bibfnamefont {A.~H.}\ \bibnamefont {MacDonald}},\ }\bibfield  {title} {\enquote {\bibinfo {title} {Topological {Insulators} in {Twisted} {Transition} {Metal} {Dichalcogenide} {Homobilayers}},}\ }\href {\doibase 10.1103/PhysRevLett.122.086402} {\bibfield  {journal} {\bibinfo  {journal} {Physical Review Letters}\ }\textbf {\bibinfo {volume} {122}},\ \bibinfo {pages} {086402} (\bibinfo {year} {2019})}\BibitemShut {NoStop}%
\bibitem [{\citenamefont {Devakul}\ \emph {et~al.}(2021)\citenamefont {Devakul}, \citenamefont {Cr{\'e}pel}, \citenamefont {Zhang},\ and\ \citenamefont {Fu}}]{devakul2021magic}%
  \BibitemOpen
  \bibfield  {author} {\bibinfo {author} {\bibfnamefont {Trithep}\ \bibnamefont {Devakul}}, \bibinfo {author} {\bibfnamefont {Valentin}\ \bibnamefont {Cr{\'e}pel}}, \bibinfo {author} {\bibfnamefont {Yang}\ \bibnamefont {Zhang}}, \ and\ \bibinfo {author} {\bibfnamefont {Liang}\ \bibnamefont {Fu}},\ }\bibfield  {title} {\enquote {\bibinfo {title} {Magic in twisted transition metal dichalcogenide bilayers},}\ }\href {https://doi.org/10.1038/s41467-021-27042-9} {\bibfield  {journal} {\bibinfo  {journal} {Nature communications}\ }\textbf {\bibinfo {volume} {12}},\ \bibinfo {pages} {6730} (\bibinfo {year} {2021})}\BibitemShut {NoStop}%
\bibitem [{\citenamefont {Kang}\ \emph {et~al.}(2024{\natexlab{a}})\citenamefont {Kang}, \citenamefont {Shen}, \citenamefont {Qiu}, \citenamefont {Zeng}, \citenamefont {Xia}, \citenamefont {Watanabe}, \citenamefont {Taniguchi}, \citenamefont {Shan},\ and\ \citenamefont {Mak}}]{kang_evidence_2024}%
  \BibitemOpen
  \bibfield  {author} {\bibinfo {author} {\bibfnamefont {Kaifei}\ \bibnamefont {Kang}}, \bibinfo {author} {\bibfnamefont {Bowen}\ \bibnamefont {Shen}}, \bibinfo {author} {\bibfnamefont {Yichen}\ \bibnamefont {Qiu}}, \bibinfo {author} {\bibfnamefont {Yihang}\ \bibnamefont {Zeng}}, \bibinfo {author} {\bibfnamefont {Zhengchao}\ \bibnamefont {Xia}}, \bibinfo {author} {\bibfnamefont {Kenji}\ \bibnamefont {Watanabe}}, \bibinfo {author} {\bibfnamefont {Takashi}\ \bibnamefont {Taniguchi}}, \bibinfo {author} {\bibfnamefont {Jie}\ \bibnamefont {Shan}}, \ and\ \bibinfo {author} {\bibfnamefont {Kin~Fai}\ \bibnamefont {Mak}},\ }\bibfield  {title} {\enquote {\bibinfo {title} {Evidence of the fractional quantum spin {Hall} effect in moiré {MoTe2}},}\ }\href {\doibase 10.1038/s41586-024-07214-5} {\bibfield  {journal} {\bibinfo  {journal} {Nature}\ }\textbf {\bibinfo {volume} {628}},\ \bibinfo {pages} {522--526} (\bibinfo {year} {2024}{\natexlab{a}})}\BibitemShut {NoStop}%
\bibitem [{\citenamefont {Foutty}\ \emph {et~al.}(2024)\citenamefont {Foutty}, \citenamefont {Kometter}, \citenamefont {Devakul}, \citenamefont {Reddy}, \citenamefont {Watanabe}, \citenamefont {Taniguchi}, \citenamefont {Fu},\ and\ \citenamefont {Feldman}}]{fouttyMappingTwisttunedMultiband2024}%
  \BibitemOpen
  \bibfield  {author} {\bibinfo {author} {\bibfnamefont {Benjamin~A.}\ \bibnamefont {Foutty}}, \bibinfo {author} {\bibfnamefont {Carlos~R.}\ \bibnamefont {Kometter}}, \bibinfo {author} {\bibfnamefont {Trithep}\ \bibnamefont {Devakul}}, \bibinfo {author} {\bibfnamefont {Aidan~P.}\ \bibnamefont {Reddy}}, \bibinfo {author} {\bibfnamefont {Kenji}\ \bibnamefont {Watanabe}}, \bibinfo {author} {\bibfnamefont {Takashi}\ \bibnamefont {Taniguchi}}, \bibinfo {author} {\bibfnamefont {Liang}\ \bibnamefont {Fu}}, \ and\ \bibinfo {author} {\bibfnamefont {Benjamin~E.}\ \bibnamefont {Feldman}},\ }\bibfield  {title} {\enquote {\bibinfo {title} {Mapping twist-tuned multiband topology in bilayer {{WSe2}}},}\ }\href {\doibase 10.1126/science.adi4728} {\bibfield  {journal} {\bibinfo  {journal} {Science}\ }\textbf {\bibinfo {volume} {384}},\ \bibinfo {pages} {343--347} (\bibinfo {year} {2024})}\BibitemShut {NoStop}%
\bibitem [{\citenamefont {Kang}\ \emph {et~al.}(2024{\natexlab{b}})\citenamefont {Kang}, \citenamefont {Qiu}, \citenamefont {Watanabe}, \citenamefont {Taniguchi}, \citenamefont {Shan},\ and\ \citenamefont {Mak}}]{kang2024observationWSe2}%
  \BibitemOpen
  \bibfield  {author} {\bibinfo {author} {\bibfnamefont {Kaifei}\ \bibnamefont {Kang}}, \bibinfo {author} {\bibfnamefont {Yichen}\ \bibnamefont {Qiu}}, \bibinfo {author} {\bibfnamefont {Kenji}\ \bibnamefont {Watanabe}}, \bibinfo {author} {\bibfnamefont {Takashi}\ \bibnamefont {Taniguchi}}, \bibinfo {author} {\bibfnamefont {Jie}\ \bibnamefont {Shan}}, \ and\ \bibinfo {author} {\bibfnamefont {Kin~Fai}\ \bibnamefont {Mak}},\ }\bibfield  {title} {\enquote {\bibinfo {title} {Observation of the double quantum spin hall phase in moir$\backslash$'e wse2},}\ }\href {https://doi.org/10.48550/arXiv.2402.04196} {\bibfield  {journal} {\bibinfo  {journal} {arXiv preprint arXiv:2402.04196}\ } (\bibinfo {year} {2024}{\natexlab{b}})}\BibitemShut {NoStop}%
\bibitem [{\citenamefont {May-Mann}\ \emph {et~al.}(2024)\citenamefont {May-Mann}, \citenamefont {Stern},\ and\ \citenamefont {Devakul}}]{may2024theory}%
  \BibitemOpen
  \bibfield  {author} {\bibinfo {author} {\bibfnamefont {Julian}\ \bibnamefont {May-Mann}}, \bibinfo {author} {\bibfnamefont {Ady}\ \bibnamefont {Stern}}, \ and\ \bibinfo {author} {\bibfnamefont {Trithep}\ \bibnamefont {Devakul}},\ }\bibfield  {title} {\enquote {\bibinfo {title} {Theory of half-integer fractional quantum spin hall insulator edges},}\ }\href {https://doi.org/10.48550/arXiv.2403.03964} {\bibfield  {journal} {\bibinfo  {journal} {arXiv preprint arXiv:2403.03964}\ } (\bibinfo {year} {2024})}\BibitemShut {NoStop}%
\bibitem [{\citenamefont {Jian}\ and\ \citenamefont {Xu}(2024)}]{jian2024minimal}%
  \BibitemOpen
  \bibfield  {author} {\bibinfo {author} {\bibfnamefont {Chao-Ming}\ \bibnamefont {Jian}}\ and\ \bibinfo {author} {\bibfnamefont {Cenke}\ \bibnamefont {Xu}},\ }\bibfield  {title} {\enquote {\bibinfo {title} {Minimal fractional topological insulator in half-filled conjugate moir$\backslash$'$\{$e$\}$ chern bands},}\ }\href {https://doi.org/10.48550/arXiv.2403.07054} {\bibfield  {journal} {\bibinfo  {journal} {arXiv preprint arXiv:2403.07054}\ } (\bibinfo {year} {2024})}\BibitemShut {NoStop}%
\bibitem [{\citenamefont {Villadiego}(2024)}]{villadiego2024halperin}%
  \BibitemOpen
  \bibfield  {author} {\bibinfo {author} {\bibfnamefont {Inti~Sodemann}\ \bibnamefont {Villadiego}},\ }\bibfield  {title} {\enquote {\bibinfo {title} {Halperin states of particles and holes in ideal time reversal invariant pairs of chern bands and the fractional quantum spin hall effect in moir$\backslash$'e mote $ \_2$},}\ }\href {https://doi.org/10.48550/arXiv.2403.12185} {\bibfield  {journal} {\bibinfo  {journal} {arXiv preprint arXiv:2403.12185}\ } (\bibinfo {year} {2024})}\BibitemShut {NoStop}%
\bibitem [{\citenamefont {Zhang}(2024)}]{zhang2024non}%
  \BibitemOpen
  \bibfield  {author} {\bibinfo {author} {\bibfnamefont {Ya-Hui}\ \bibnamefont {Zhang}},\ }\bibfield  {title} {\enquote {\bibinfo {title} {Non-abelian and abelian descendants of vortex spin liquid: fractional quantum spin hall effect in twisted mote $ \_2$},}\ }\href@noop {} {\bibfield  {journal} {\bibinfo  {journal} {arXiv preprint arXiv:2403.12126}\ } (\bibinfo {year} {2024})}\BibitemShut {NoStop}%
\bibitem [{\citenamefont {Chou}\ and\ \citenamefont {Sarma}(2024)}]{chou2024composite}%
  \BibitemOpen
  \bibfield  {author} {\bibinfo {author} {\bibfnamefont {Yang-Zhi}\ \bibnamefont {Chou}}\ and\ \bibinfo {author} {\bibfnamefont {Sankar~Das}\ \bibnamefont {Sarma}},\ }\bibfield  {title} {\enquote {\bibinfo {title} {Composite helical edges from abelian fractional topological insulators},}\ }\href {https://arxiv.org/abs/2406.06669} {\bibfield  {journal} {\bibinfo  {journal} {arXiv preprint arXiv:2406.06669}\ } (\bibinfo {year} {2024})}\BibitemShut {NoStop}%
\bibitem [{\citenamefont {Reddy}\ \emph {et~al.}(2024)\citenamefont {Reddy}, \citenamefont {Paul}, \citenamefont {Abouelkomsan},\ and\ \citenamefont {Fu}}]{reddy2024non}%
  \BibitemOpen
  \bibfield  {author} {\bibinfo {author} {\bibfnamefont {Aidan~P}\ \bibnamefont {Reddy}}, \bibinfo {author} {\bibfnamefont {Nisarga}\ \bibnamefont {Paul}}, \bibinfo {author} {\bibfnamefont {Ahmed}\ \bibnamefont {Abouelkomsan}}, \ and\ \bibinfo {author} {\bibfnamefont {Liang}\ \bibnamefont {Fu}},\ }\bibfield  {title} {\enquote {\bibinfo {title} {Non-abelian fractionalization in topological minibands},}\ }\href {https://doi.org/10.48550/arXiv.2403.00059} {\bibfield  {journal} {\bibinfo  {journal} {arXiv preprint arXiv:2403.00059}\ } (\bibinfo {year} {2024})}\BibitemShut {NoStop}%
\bibitem [{\citenamefont {Moore}\ and\ \citenamefont {Read}(1991)}]{Moore1991Aug}%
  \BibitemOpen
  \bibfield  {author} {\bibinfo {author} {\bibfnamefont {Gregory}\ \bibnamefont {Moore}}\ and\ \bibinfo {author} {\bibfnamefont {Nicholas}\ \bibnamefont {Read}},\ }\bibfield  {title} {\enquote {\bibinfo {title} {{Nonabelions in the fractional quantum hall effect}},}\ }\href {\doibase 10.1016/0550-3213(91)90407-O} {\bibfield  {journal} {\bibinfo  {journal} {Nucl. Phys. B}\ }\textbf {\bibinfo {volume} {360}},\ \bibinfo {pages} {362--396} (\bibinfo {year} {1991})}\BibitemShut {NoStop}%
\bibitem [{\citenamefont {Read}\ and\ \citenamefont {Green}(2000)}]{Read2000Apr}%
  \BibitemOpen
  \bibfield  {author} {\bibinfo {author} {\bibfnamefont {N.}~\bibnamefont {Read}}\ and\ \bibinfo {author} {\bibfnamefont {Dmitry}\ \bibnamefont {Green}},\ }\bibfield  {title} {\enquote {\bibinfo {title} {{Paired states of fermions in two dimensions with breaking of parity and time-reversal symmetries and the fractional quantum Hall effect}},}\ }\href {\doibase 10.1103/PhysRevB.61.10267} {\bibfield  {journal} {\bibinfo  {journal} {Phys. Rev. B}\ }\textbf {\bibinfo {volume} {61}},\ \bibinfo {pages} {10267--10297} (\bibinfo {year} {2000})}\BibitemShut {NoStop}%
\bibitem [{\citenamefont {Paul}\ \emph {et~al.}(2022)\citenamefont {Paul}, \citenamefont {Crowley}, \citenamefont {Devakul},\ and\ \citenamefont {Fu}}]{paul_magic2022}%
  \BibitemOpen
  \bibfield  {author} {\bibinfo {author} {\bibfnamefont {Nisarga}\ \bibnamefont {Paul}}, \bibinfo {author} {\bibfnamefont {Philip J.~D.}\ \bibnamefont {Crowley}}, \bibinfo {author} {\bibfnamefont {Trithep}\ \bibnamefont {Devakul}}, \ and\ \bibinfo {author} {\bibfnamefont {Liang}\ \bibnamefont {Fu}},\ }\bibfield  {title} {\enquote {\bibinfo {title} {Moir\'e landau fans and magic zeros},}\ }\href {\doibase 10.1103/PhysRevLett.129.116804} {\bibfield  {journal} {\bibinfo  {journal} {Phys. Rev. Lett.}\ }\textbf {\bibinfo {volume} {129}},\ \bibinfo {pages} {116804} (\bibinfo {year} {2022})}\BibitemShut {NoStop}%
\bibitem [{\citenamefont {Morales-Dur\'an}\ \emph {et~al.}(2024)\citenamefont {Morales-Dur\'an}, \citenamefont {Wei}, \citenamefont {Shi},\ and\ \citenamefont {MacDonald}}]{morales2023magic}%
  \BibitemOpen
  \bibfield  {author} {\bibinfo {author} {\bibfnamefont {Nicol\'as}\ \bibnamefont {Morales-Dur\'an}}, \bibinfo {author} {\bibfnamefont {Nemin}\ \bibnamefont {Wei}}, \bibinfo {author} {\bibfnamefont {Jingtian}\ \bibnamefont {Shi}}, \ and\ \bibinfo {author} {\bibfnamefont {Allan~H.}\ \bibnamefont {MacDonald}},\ }\bibfield  {title} {\enquote {\bibinfo {title} {Magic angles and fractional chern insulators in twisted homobilayer transition metal dichalcogenides},}\ }\href {\doibase 10.1103/PhysRevLett.132.096602} {\bibfield  {journal} {\bibinfo  {journal} {Phys. Rev. Lett.}\ }\textbf {\bibinfo {volume} {132}},\ \bibinfo {pages} {096602} (\bibinfo {year} {2024})}\BibitemShut {NoStop}%
\bibitem [{sup()}]{supp}%
  \BibitemOpen
  \href@noop {} {}\bibinfo {note} {See Supplemental Material for details and additional results}\BibitemShut {NoStop}%
\bibitem [{\citenamefont {Girvin}\ \emph {et~al.}(1986)\citenamefont {Girvin}, \citenamefont {MacDonald},\ and\ \citenamefont {Platzman}}]{girvinMagnetorotonTheoryCollective1986}%
  \BibitemOpen
  \bibfield  {author} {\bibinfo {author} {\bibfnamefont {S.~M.}\ \bibnamefont {Girvin}}, \bibinfo {author} {\bibfnamefont {A.~H.}\ \bibnamefont {MacDonald}}, \ and\ \bibinfo {author} {\bibfnamefont {P.~M.}\ \bibnamefont {Platzman}},\ }\bibfield  {title} {\enquote {\bibinfo {title} {Magneto-roton theory of collective excitations in the fractional quantum {{Hall}} effect},}\ }\href {\doibase 10.1103/PhysRevB.33.2481} {\bibfield  {journal} {\bibinfo  {journal} {Physical Review B}\ }\textbf {\bibinfo {volume} {33}},\ \bibinfo {pages} {2481--2494} (\bibinfo {year} {1986})}\BibitemShut {NoStop}%
\bibitem [{\citenamefont {Haldane}(1983)}]{haldane1983fractional}%
  \BibitemOpen
  \bibfield  {author} {\bibinfo {author} {\bibfnamefont {F~Duncan~M}\ \bibnamefont {Haldane}},\ }\bibfield  {title} {\enquote {\bibinfo {title} {Fractional quantization of the hall effect: a hierarchy of incompressible quantum fluid states},}\ }\href {https://doi.org/10.1103/PhysRevLett.51.605} {\bibfield  {journal} {\bibinfo  {journal} {Physical Review Letters}\ }\textbf {\bibinfo {volume} {51}},\ \bibinfo {pages} {605} (\bibinfo {year} {1983})}\BibitemShut {NoStop}%
\bibitem [{\citenamefont {Stefanidis}\ and\ \citenamefont {Sodemann}(2020)}]{stefanidis_excitonic_2020}%
  \BibitemOpen
  \bibfield  {author} {\bibinfo {author} {\bibfnamefont {Nikolaos}\ \bibnamefont {Stefanidis}}\ and\ \bibinfo {author} {\bibfnamefont {Inti}\ \bibnamefont {Sodemann}},\ }\bibfield  {title} {\enquote {\bibinfo {title} {Excitonic {Laughlin} {States} in {Ideal} {Topological} {Insulator} {Flat} {Bands} and {Possible} {Presence} in {Moir}{\textbackslash}'e {Superlattice} {Materials}},}\ }\href {\doibase 10.1103/PhysRevB.102.035158} {\bibfield  {journal} {\bibinfo  {journal} {Physical Review B}\ }\textbf {\bibinfo {volume} {102}},\ \bibinfo {pages} {035158} (\bibinfo {year} {2020})}\BibitemShut {NoStop}%
\bibitem [{\citenamefont {Gor’kov}\ and\ \citenamefont {Dzyaloshinskii}(1968)}]{gor1968contribution}%
  \BibitemOpen
  \bibfield  {author} {\bibinfo {author} {\bibfnamefont {LP}~\bibnamefont {Gor’kov}}\ and\ \bibinfo {author} {\bibfnamefont {IE}~\bibnamefont {Dzyaloshinskii}},\ }\bibfield  {title} {\enquote {\bibinfo {title} {Contribution to the theory of the mott exciton in a strong magnetic field},}\ }\href@noop {} {\bibfield  {journal} {\bibinfo  {journal} {Sov. Phys. JETP}\ }\textbf {\bibinfo {volume} {26}},\ \bibinfo {pages} {449--451} (\bibinfo {year} {1968})}\BibitemShut {NoStop}%
\bibitem [{\citenamefont {Kallin}\ and\ \citenamefont {Halperin}(1984)}]{kallin_excitations_1984}%
  \BibitemOpen
  \bibfield  {author} {\bibinfo {author} {\bibfnamefont {C.}~\bibnamefont {Kallin}}\ and\ \bibinfo {author} {\bibfnamefont {B.~I.}\ \bibnamefont {Halperin}},\ }\bibfield  {title} {\enquote {\bibinfo {title} {Excitations from a filled {Landau} level in the two-dimensional electron gas},}\ }\href {\doibase 10.1103/PhysRevB.30.5655} {\bibfield  {journal} {\bibinfo  {journal} {Physical Review B}\ }\textbf {\bibinfo {volume} {30}},\ \bibinfo {pages} {5655--5668} (\bibinfo {year} {1984})}\BibitemShut {NoStop}%
\bibitem [{\citenamefont {Sodemann}\ \emph {et~al.}(2017)\citenamefont {Sodemann}, \citenamefont {Zhu},\ and\ \citenamefont {Fu}}]{sodemann_quantum_2017}%
  \BibitemOpen
  \bibfield  {author} {\bibinfo {author} {\bibfnamefont {Inti}\ \bibnamefont {Sodemann}}, \bibinfo {author} {\bibfnamefont {Zheng}\ \bibnamefont {Zhu}}, \ and\ \bibinfo {author} {\bibfnamefont {Liang}\ \bibnamefont {Fu}},\ }\bibfield  {title} {\enquote {\bibinfo {title} {Quantum {Hall} {Ferroelectrics} and {Nematics} in {Multivalley} {Systems}},}\ }\href {\doibase 10.1103/PhysRevX.7.041068} {\bibfield  {journal} {\bibinfo  {journal} {Physical Review X}\ }\textbf {\bibinfo {volume} {7}},\ \bibinfo {pages} {041068} (\bibinfo {year} {2017})}\BibitemShut {NoStop}%
\bibitem [{\citenamefont {Bultinck}\ \emph {et~al.}(2020{\natexlab{a}})\citenamefont {Bultinck}, \citenamefont {Khalaf}, \citenamefont {Liu}, \citenamefont {Chatterjee}, \citenamefont {Vishwanath},\ and\ \citenamefont {Zaletel}}]{bultinck_ground_2020}%
  \BibitemOpen
  \bibfield  {author} {\bibinfo {author} {\bibfnamefont {Nick}\ \bibnamefont {Bultinck}}, \bibinfo {author} {\bibfnamefont {Eslam}\ \bibnamefont {Khalaf}}, \bibinfo {author} {\bibfnamefont {Shang}\ \bibnamefont {Liu}}, \bibinfo {author} {\bibfnamefont {Shubhayu}\ \bibnamefont {Chatterjee}}, \bibinfo {author} {\bibfnamefont {Ashvin}\ \bibnamefont {Vishwanath}}, \ and\ \bibinfo {author} {\bibfnamefont {Michael~P.}\ \bibnamefont {Zaletel}},\ }\bibfield  {title} {\enquote {\bibinfo {title} {Ground {State} and {Hidden} {Symmetry} of {Magic}-{Angle} {Graphene} at {Even} {Integer} {Filling}},}\ }\href {\doibase 10.1103/PhysRevX.10.031034} {\bibfield  {journal} {\bibinfo  {journal} {Physical Review X}\ }\textbf {\bibinfo {volume} {10}},\ \bibinfo {pages} {031034} (\bibinfo {year} {2020}{\natexlab{a}})}\BibitemShut {NoStop}%
\bibitem [{\citenamefont {Bultinck}\ \emph {et~al.}(2020{\natexlab{b}})\citenamefont {Bultinck}, \citenamefont {Chatterjee},\ and\ \citenamefont {Zaletel}}]{bultinck_mechanism_2020}%
  \BibitemOpen
  \bibfield  {author} {\bibinfo {author} {\bibfnamefont {Nick}\ \bibnamefont {Bultinck}}, \bibinfo {author} {\bibfnamefont {Shubhayu}\ \bibnamefont {Chatterjee}}, \ and\ \bibinfo {author} {\bibfnamefont {Michael~P.}\ \bibnamefont {Zaletel}},\ }\bibfield  {title} {\enquote {\bibinfo {title} {Mechanism for {Anomalous} {Hall} {Ferromagnetism} in {Twisted} {Bilayer} {Graphene}},}\ }\href {\doibase 10.1103/PhysRevLett.124.166601} {\bibfield  {journal} {\bibinfo  {journal} {Physical Review Letters}\ }\textbf {\bibinfo {volume} {124}},\ \bibinfo {pages} {166601} (\bibinfo {year} {2020}{\natexlab{b}})},\ \bibinfo {note} {publisher: American Physical Society}\BibitemShut {NoStop}%
\bibitem [{\citenamefont {Lian}\ \emph {et~al.}(2021)\citenamefont {Lian}, \citenamefont {Song}, \citenamefont {Regnault}, \citenamefont {Efetov}, \citenamefont {Yazdani},\ and\ \citenamefont {Bernevig}}]{lian_twisted_2021}%
  \BibitemOpen
  \bibfield  {author} {\bibinfo {author} {\bibfnamefont {Biao}\ \bibnamefont {Lian}}, \bibinfo {author} {\bibfnamefont {Zhi-Da}\ \bibnamefont {Song}}, \bibinfo {author} {\bibfnamefont {Nicolas}\ \bibnamefont {Regnault}}, \bibinfo {author} {\bibfnamefont {Dmitri~K.}\ \bibnamefont {Efetov}}, \bibinfo {author} {\bibfnamefont {Ali}\ \bibnamefont {Yazdani}}, \ and\ \bibinfo {author} {\bibfnamefont {B.~Andrei}\ \bibnamefont {Bernevig}},\ }\bibfield  {title} {\enquote {\bibinfo {title} {Twisted bilayer graphene. {IV}. {Exact} insulator ground states and phase diagram},}\ }\href {\doibase 10.1103/PhysRevB.103.205414} {\bibfield  {journal} {\bibinfo  {journal} {Physical Review B}\ }\textbf {\bibinfo {volume} {103}},\ \bibinfo {pages} {205414} (\bibinfo {year} {2021})}\BibitemShut {NoStop}%
\bibitem [{\citenamefont {Repellin}\ \emph {et~al.}(2020)\citenamefont {Repellin}, \citenamefont {Dong}, \citenamefont {Zhang},\ and\ \citenamefont {Senthil}}]{repellin_ferromagnetism_2020}%
  \BibitemOpen
  \bibfield  {author} {\bibinfo {author} {\bibfnamefont {Cécile}\ \bibnamefont {Repellin}}, \bibinfo {author} {\bibfnamefont {Zhihuan}\ \bibnamefont {Dong}}, \bibinfo {author} {\bibfnamefont {Ya-Hui}\ \bibnamefont {Zhang}}, \ and\ \bibinfo {author} {\bibfnamefont {T.}~\bibnamefont {Senthil}},\ }\bibfield  {title} {\enquote {\bibinfo {title} {Ferromagnetism in {Narrow} {Bands} of {Moir}{\textbackslash}'e {Superlattices}},}\ }\href {\doibase 10.1103/PhysRevLett.124.187601} {\bibfield  {journal} {\bibinfo  {journal} {Physical Review Letters}\ }\textbf {\bibinfo {volume} {124}},\ \bibinfo {pages} {187601} (\bibinfo {year} {2020})}\BibitemShut {NoStop}%
\bibitem [{\citenamefont {Kwan}\ \emph {et~al.}(2021)\citenamefont {Kwan}, \citenamefont {Hu}, \citenamefont {Simon},\ and\ \citenamefont {Parameswaran}}]{kwan_exciton_2021}%
  \BibitemOpen
  \bibfield  {author} {\bibinfo {author} {\bibfnamefont {Yves~H.}\ \bibnamefont {Kwan}}, \bibinfo {author} {\bibfnamefont {Yichen}\ \bibnamefont {Hu}}, \bibinfo {author} {\bibfnamefont {Steven~H.}\ \bibnamefont {Simon}}, \ and\ \bibinfo {author} {\bibfnamefont {S.~A.}\ \bibnamefont {Parameswaran}},\ }\bibfield  {title} {\enquote {\bibinfo {title} {Exciton {Band} {Topology} in {Spontaneous} {Quantum} {Anomalous} {Hall} {Insulators}: {Applications} to {Twisted} {Bilayer} {Graphene}},}\ }\href {\doibase 10.1103/PhysRevLett.126.137601} {\bibfield  {journal} {\bibinfo  {journal} {Physical Review Letters}\ }\textbf {\bibinfo {volume} {126}},\ \bibinfo {pages} {137601} (\bibinfo {year} {2021})}\BibitemShut {NoStop}%
\bibitem [{\citenamefont {Kwan}\ \emph {et~al.}(2022)\citenamefont {Kwan}, \citenamefont {Hu}, \citenamefont {Simon},\ and\ \citenamefont {Parameswaran}}]{kwan_excitonic_2022}%
  \BibitemOpen
  \bibfield  {author} {\bibinfo {author} {\bibfnamefont {Yves~H.}\ \bibnamefont {Kwan}}, \bibinfo {author} {\bibfnamefont {Yichen}\ \bibnamefont {Hu}}, \bibinfo {author} {\bibfnamefont {Steven~H.}\ \bibnamefont {Simon}}, \ and\ \bibinfo {author} {\bibfnamefont {S.~A.}\ \bibnamefont {Parameswaran}},\ }\bibfield  {title} {\enquote {\bibinfo {title} {Excitonic fractional quantum {Hall} hierarchy in moir{\textbackslash}'e heterostructures},}\ }\href {\doibase 10.1103/PhysRevB.105.235121} {\bibfield  {journal} {\bibinfo  {journal} {Physical Review B}\ }\textbf {\bibinfo {volume} {105}},\ \bibinfo {pages} {235121} (\bibinfo {year} {2022})}\BibitemShut {NoStop}%
\bibitem [{\citenamefont {Abouelkomsan}\ \emph {et~al.}(2020)\citenamefont {Abouelkomsan}, \citenamefont {Liu},\ and\ \citenamefont {Bergholtz}}]{abouelkomsan2020particle}%
  \BibitemOpen
  \bibfield  {author} {\bibinfo {author} {\bibfnamefont {Ahmed}\ \bibnamefont {Abouelkomsan}}, \bibinfo {author} {\bibfnamefont {Zhao}\ \bibnamefont {Liu}}, \ and\ \bibinfo {author} {\bibfnamefont {Emil~J}\ \bibnamefont {Bergholtz}},\ }\bibfield  {title} {\enquote {\bibinfo {title} {Particle-hole duality, emergent fermi liquids, and fractional chern insulators in moir{\'e} flatbands},}\ }\href {\doibase 10.1103/PhysRevLett.124.106803} {\bibfield  {journal} {\bibinfo  {journal} {Physical review letters}\ }\textbf {\bibinfo {volume} {124}},\ \bibinfo {pages} {106803} (\bibinfo {year} {2020})}\BibitemShut {NoStop}%
\bibitem [{Note1()}]{Note1}%
  \BibitemOpen
  \bibinfo {note} {Due to our choice of interactions, our problem can be exactly mapped to a quantum Hall bilayer in opposite magnetic fields therefore this spin-dependent constant term naturally arises due to the charge imbalance of both layers.}\BibitemShut {Stop}%
\bibitem [{\citenamefont {MacDonald}\ \emph {et~al.}(1990)\citenamefont {MacDonald}, \citenamefont {Platzman},\ and\ \citenamefont {Boebinger}}]{macdonald_collapse_1990}%
  \BibitemOpen
  \bibfield  {author} {\bibinfo {author} {\bibfnamefont {A.~H.}\ \bibnamefont {MacDonald}}, \bibinfo {author} {\bibfnamefont {P.~M.}\ \bibnamefont {Platzman}}, \ and\ \bibinfo {author} {\bibfnamefont {G.~S.}\ \bibnamefont {Boebinger}},\ }\bibfield  {title} {\enquote {\bibinfo {title} {Collapse of integer {Hall} gaps in a double-quantum-well system},}\ }\href {\doibase 10.1103/PhysRevLett.65.775} {\bibfield  {journal} {\bibinfo  {journal} {Physical Review Letters}\ }\textbf {\bibinfo {volume} {65}},\ \bibinfo {pages} {775--778} (\bibinfo {year} {1990})}\BibitemShut {NoStop}%
\bibitem [{\citenamefont {Zhu}\ \emph {et~al.}(2019)\citenamefont {Zhu}, \citenamefont {Jian},\ and\ \citenamefont {Sheng}}]{zhu_exciton_2019}%
  \BibitemOpen
  \bibfield  {author} {\bibinfo {author} {\bibfnamefont {Zheng}\ \bibnamefont {Zhu}}, \bibinfo {author} {\bibfnamefont {Shao-Kai}\ \bibnamefont {Jian}}, \ and\ \bibinfo {author} {\bibfnamefont {D.~N.}\ \bibnamefont {Sheng}},\ }\bibfield  {title} {\enquote {\bibinfo {title} {Exciton condensation in quantum {Hall} bilayers at total filling $\nu_t =5$},}\ }\href {\doibase 10.1103/PhysRevB.99.201108} {\bibfield  {journal} {\bibinfo  {journal} {Physical Review B}\ }\textbf {\bibinfo {volume} {99}},\ \bibinfo {pages} {201108} (\bibinfo {year} {2019})}\BibitemShut {NoStop}%
\bibitem [{\citenamefont {Zhu}\ \emph {et~al.}(2017)\citenamefont {Zhu}, \citenamefont {Fu},\ and\ \citenamefont {Sheng}}]{zhu_numerical_2017}%
  \BibitemOpen
  \bibfield  {author} {\bibinfo {author} {\bibfnamefont {Zheng}\ \bibnamefont {Zhu}}, \bibinfo {author} {\bibfnamefont {Liang}\ \bibnamefont {Fu}}, \ and\ \bibinfo {author} {\bibfnamefont {D.~N.}\ \bibnamefont {Sheng}},\ }\bibfield  {title} {\enquote {\bibinfo {title} {Numerical {Study} of {Quantum} {Hall} {Bilayers} at {Total} {Filling} $\nu_t = 1$: {A} {New} {Phase} at {Intermediate} {Layer} {Distances}},}\ }\href {\doibase 10.1103/PhysRevLett.119.177601} {\bibfield  {journal} {\bibinfo  {journal} {Physical Review Letters}\ }\textbf {\bibinfo {volume} {119}},\ \bibinfo {pages} {177601} (\bibinfo {year} {2017})},\ \bibinfo {note} {publisher: American Physical Society}\BibitemShut {NoStop}%
\bibitem [{\citenamefont {Ardonne}\ \emph {et~al.}(2008)\citenamefont {Ardonne}, \citenamefont {Bergholtz}, \citenamefont {Kailasvuori},\ and\ \citenamefont {Wikberg}}]{ardonne2008degeneracy}%
  \BibitemOpen
  \bibfield  {author} {\bibinfo {author} {\bibfnamefont {Eddy}\ \bibnamefont {Ardonne}}, \bibinfo {author} {\bibfnamefont {Emil~J}\ \bibnamefont {Bergholtz}}, \bibinfo {author} {\bibfnamefont {Janik}\ \bibnamefont {Kailasvuori}}, \ and\ \bibinfo {author} {\bibfnamefont {Emma}\ \bibnamefont {Wikberg}},\ }\bibfield  {title} {\enquote {\bibinfo {title} {Degeneracy of non-abelian quantum hall states on the torus: domain walls and conformal field theory},}\ }\href {https://iopscience.iop.org/article/10.1088/1742-5468/2008/04/P04016} {\bibfield  {journal} {\bibinfo  {journal} {Journal of Statistical Mechanics: Theory and Experiment}\ }\textbf {\bibinfo {volume} {2008}},\ \bibinfo {pages} {P04016} (\bibinfo {year} {2008})}\BibitemShut {NoStop}%
\bibitem [{\citenamefont {Oshikawa}\ \emph {et~al.}(2007)\citenamefont {Oshikawa}, \citenamefont {Kim}, \citenamefont {Shtengel}, \citenamefont {Nayak},\ and\ \citenamefont {Tewari}}]{Oshikawa2007Jun}%
  \BibitemOpen
  \bibfield  {author} {\bibinfo {author} {\bibfnamefont {Masaki}\ \bibnamefont {Oshikawa}}, \bibinfo {author} {\bibfnamefont {Yong~Baek}\ \bibnamefont {Kim}}, \bibinfo {author} {\bibfnamefont {Kirill}\ \bibnamefont {Shtengel}}, \bibinfo {author} {\bibfnamefont {Chetan}\ \bibnamefont {Nayak}}, \ and\ \bibinfo {author} {\bibfnamefont {Sumanta}\ \bibnamefont {Tewari}},\ }\bibfield  {title} {\enquote {\bibinfo {title} {{Topological degeneracy of non-Abelian states for dummies}},}\ }\href {\doibase 10.1016/j.aop.2006.08.001} {\bibfield  {journal} {\bibinfo  {journal} {Ann. Phys.}\ }\textbf {\bibinfo {volume} {322}},\ \bibinfo {pages} {1477--1498} (\bibinfo {year} {2007})}\BibitemShut {NoStop}%
\bibitem [{\citenamefont {Lee}\ \emph {et~al.}(2007)\citenamefont {Lee}, \citenamefont {Ryu}, \citenamefont {Nayak},\ and\ \citenamefont {Fisher}}]{lee2007particle}%
  \BibitemOpen
  \bibfield  {author} {\bibinfo {author} {\bibfnamefont {Sung-Sik}\ \bibnamefont {Lee}}, \bibinfo {author} {\bibfnamefont {Shinsei}\ \bibnamefont {Ryu}}, \bibinfo {author} {\bibfnamefont {Chetan}\ \bibnamefont {Nayak}}, \ and\ \bibinfo {author} {\bibfnamefont {Matthew~PA}\ \bibnamefont {Fisher}},\ }\bibfield  {title} {\enquote {\bibinfo {title} {Particle-hole symmetry and the $\nu$= 5 2 quantum hall state},}\ }\href {https://doi.org/10.1103/PhysRevLett.99.236807} {\bibfield  {journal} {\bibinfo  {journal} {Physical review letters}\ }\textbf {\bibinfo {volume} {99}},\ \bibinfo {pages} {236807} (\bibinfo {year} {2007})}\BibitemShut {NoStop}%
\bibitem [{\citenamefont {Levin}\ \emph {et~al.}(2007)\citenamefont {Levin}, \citenamefont {Halperin},\ and\ \citenamefont {Rosenow}}]{levin2007particle}%
  \BibitemOpen
  \bibfield  {author} {\bibinfo {author} {\bibfnamefont {Michael}\ \bibnamefont {Levin}}, \bibinfo {author} {\bibfnamefont {Bertrand~I}\ \bibnamefont {Halperin}}, \ and\ \bibinfo {author} {\bibfnamefont {Bernd}\ \bibnamefont {Rosenow}},\ }\bibfield  {title} {\enquote {\bibinfo {title} {Particle-hole symmetry and the pfaffian state},}\ }\href {https://doi.org/10.1103/PhysRevLett.99.236806} {\bibfield  {journal} {\bibinfo  {journal} {Physical review letters}\ }\textbf {\bibinfo {volume} {99}},\ \bibinfo {pages} {236806} (\bibinfo {year} {2007})}\BibitemShut {NoStop}%
\bibitem [{\citenamefont {Read}\ and\ \citenamefont {Rezayi}(1999)}]{Read1999Mar}%
  \BibitemOpen
  \bibfield  {author} {\bibinfo {author} {\bibfnamefont {N.}~\bibnamefont {Read}}\ and\ \bibinfo {author} {\bibfnamefont {E.}~\bibnamefont {Rezayi}},\ }\bibfield  {title} {\enquote {\bibinfo {title} {{Beyond paired quantum Hall states: Parafermions and incompressible states in the first excited Landau level}},}\ }\href {\doibase 10.1103/PhysRevB.59.8084} {\bibfield  {journal} {\bibinfo  {journal} {Phys. Rev. B}\ }\textbf {\bibinfo {volume} {59}},\ \bibinfo {pages} {8084--8092} (\bibinfo {year} {1999})}\BibitemShut {NoStop}%
\bibitem [{\citenamefont {Greiter}\ \emph {et~al.}(1991)\citenamefont {Greiter}, \citenamefont {Wen},\ and\ \citenamefont {Wilczek}}]{greiter1991paired}%
  \BibitemOpen
  \bibfield  {author} {\bibinfo {author} {\bibfnamefont {Martin}\ \bibnamefont {Greiter}}, \bibinfo {author} {\bibfnamefont {Xiao-Gang}\ \bibnamefont {Wen}}, \ and\ \bibinfo {author} {\bibfnamefont {Frank}\ \bibnamefont {Wilczek}},\ }\bibfield  {title} {\enquote {\bibinfo {title} {Paired hall state at half filling},}\ }\href@noop {} {\bibfield  {journal} {\bibinfo  {journal} {Physical review letters}\ }\textbf {\bibinfo {volume} {66}},\ \bibinfo {pages} {3205} (\bibinfo {year} {1991})}\BibitemShut {NoStop}%
\bibitem [{\citenamefont {Wen}(1993)}]{wenTopologicalOrderEdge1993}%
  \BibitemOpen
  \bibfield  {author} {\bibinfo {author} {\bibfnamefont {Xiao-Gang}\ \bibnamefont {Wen}},\ }\bibfield  {title} {\enquote {\bibinfo {title} {Topological order and edge structure of {\textbackslash}ensuremath\{{\textbackslash}nu\}=1/2 quantum {{Hall}} state},}\ }\href {\doibase 10.1103/PhysRevLett.70.355} {\bibfield  {journal} {\bibinfo  {journal} {Physical Review Letters}\ }\textbf {\bibinfo {volume} {70}},\ \bibinfo {pages} {355--358} (\bibinfo {year} {1993})}\BibitemShut {NoStop}%
\bibitem [{\citenamefont {Wu}\ \emph {et~al.}(2024)\citenamefont {Wu}, \citenamefont {Shaffer}, \citenamefont {Wu},\ and\ \citenamefont {Santos}}]{wuTimereversalInvariantTopological2024}%
  \BibitemOpen
  \bibfield  {author} {\bibinfo {author} {\bibfnamefont {Yi-Ming}\ \bibnamefont {Wu}}, \bibinfo {author} {\bibfnamefont {Daniel}\ \bibnamefont {Shaffer}}, \bibinfo {author} {\bibfnamefont {Zhengzhi}\ \bibnamefont {Wu}}, \ and\ \bibinfo {author} {\bibfnamefont {Luiz~H.}\ \bibnamefont {Santos}},\ }\bibfield  {title} {\enquote {\bibinfo {title} {Time-reversal invariant topological moir{\textbackslash}'e flat band: {{A}} platform for the fractional quantum spin {{Hall}} effect},}\ }\href {\doibase 10.1103/PhysRevB.109.115111} {\bibfield  {journal} {\bibinfo  {journal} {Physical Review B}\ }\textbf {\bibinfo {volume} {109}},\ \bibinfo {pages} {115111} (\bibinfo {year} {2024})}\BibitemShut {NoStop}%
\bibitem [{\citenamefont {Kane}\ and\ \citenamefont {Fisher}(1997)}]{kaneQuantizedThermalTransport1997a}%
  \BibitemOpen
  \bibfield  {author} {\bibinfo {author} {\bibfnamefont {C.~L.}\ \bibnamefont {Kane}}\ and\ \bibinfo {author} {\bibfnamefont {Matthew P.~A.}\ \bibnamefont {Fisher}},\ }\bibfield  {title} {\enquote {\bibinfo {title} {Quantized thermal transport in the fractional quantum {{Hall}} effect},}\ }\href {\doibase 10.1103/PhysRevB.55.15832} {\bibfield  {journal} {\bibinfo  {journal} {Physical Review B}\ }\textbf {\bibinfo {volume} {55}},\ \bibinfo {pages} {15832--15837} (\bibinfo {year} {1997})}\BibitemShut {NoStop}%
\bibitem [{\citenamefont {Cappelli}\ \emph {et~al.}(2002)\citenamefont {Cappelli}, \citenamefont {Huerta},\ and\ \citenamefont {Zemba}}]{cappelliThermalTransportChiral2002}%
  \BibitemOpen
  \bibfield  {author} {\bibinfo {author} {\bibfnamefont {Andrea}\ \bibnamefont {Cappelli}}, \bibinfo {author} {\bibfnamefont {Marina}\ \bibnamefont {Huerta}}, \ and\ \bibinfo {author} {\bibfnamefont {Guillermo~R.}\ \bibnamefont {Zemba}},\ }\bibfield  {title} {\enquote {\bibinfo {title} {Thermal transport in chiral conformal theories and hierarchical quantum {{Hall}} states},}\ }\href {\doibase 10.1016/S0550-3213(02)00340-1} {\bibfield  {journal} {\bibinfo  {journal} {Nuclear Physics B}\ }\textbf {\bibinfo {volume} {636}},\ \bibinfo {pages} {568--582} (\bibinfo {year} {2002})}\BibitemShut {NoStop}%
\bibitem [{\citenamefont {Sodemann}\ and\ \citenamefont {MacDonald}(2013)}]{sodemann2013landau}%
  \BibitemOpen
  \bibfield  {author} {\bibinfo {author} {\bibfnamefont {Inti}\ \bibnamefont {Sodemann}}\ and\ \bibinfo {author} {\bibfnamefont {AH}~\bibnamefont {MacDonald}},\ }\bibfield  {title} {\enquote {\bibinfo {title} {Landau level mixing and the fractional quantum hall effect},}\ }\href@noop {} {\bibfield  {journal} {\bibinfo  {journal} {Physical Review B}\ }\textbf {\bibinfo {volume} {87}},\ \bibinfo {pages} {245425} (\bibinfo {year} {2013})}\BibitemShut {NoStop}%
\bibitem [{\citenamefont {Simon}\ and\ \citenamefont {Rezayi}(2013)}]{simon_landau_2013}%
  \BibitemOpen
  \bibfield  {author} {\bibinfo {author} {\bibfnamefont {Steven~H.}\ \bibnamefont {Simon}}\ and\ \bibinfo {author} {\bibfnamefont {Edward~H.}\ \bibnamefont {Rezayi}},\ }\bibfield  {title} {\enquote {\bibinfo {title} {Landau {Level} {Mixing} in the {Perturbative} {Limit}},}\ }\href {\doibase 10.1103/PhysRevB.87.155426} {\bibfield  {journal} {\bibinfo  {journal} {Physical Review B}\ }\textbf {\bibinfo {volume} {87}},\ \bibinfo {pages} {155426} (\bibinfo {year} {2013})}\BibitemShut {NoStop}%
\bibitem [{\citenamefont {Rezayi}(2017)}]{rezayiLandauLevelMixing2017}%
  \BibitemOpen
  \bibfield  {author} {\bibinfo {author} {\bibfnamefont {Edward~H}\ \bibnamefont {Rezayi}},\ }\bibfield  {title} {\enquote {\bibinfo {title} {Landau level mixing and the ground state of the $\nu$= 5/2 quantum hall effect},}\ }\href {https://doi.org/10.1103/PhysRevLett.119.026801} {\bibfield  {journal} {\bibinfo  {journal} {Physical Review Letters}\ }\textbf {\bibinfo {volume} {119}},\ \bibinfo {pages} {026801} (\bibinfo {year} {2017})}\BibitemShut {NoStop}%
\bibitem [{\citenamefont {Reddy}\ \emph {et~al.}(2023)\citenamefont {Reddy}, \citenamefont {Alsallom}, \citenamefont {Zhang}, \citenamefont {Devakul},\ and\ \citenamefont {Fu}}]{reddy2023fractional}%
  \BibitemOpen
  \bibfield  {author} {\bibinfo {author} {\bibfnamefont {Aidan~P.}\ \bibnamefont {Reddy}}, \bibinfo {author} {\bibfnamefont {Faisal}\ \bibnamefont {Alsallom}}, \bibinfo {author} {\bibfnamefont {Yang}\ \bibnamefont {Zhang}}, \bibinfo {author} {\bibfnamefont {Trithep}\ \bibnamefont {Devakul}}, \ and\ \bibinfo {author} {\bibfnamefont {Liang}\ \bibnamefont {Fu}},\ }\bibfield  {title} {\enquote {\bibinfo {title} {Fractional quantum anomalous hall states in twisted bilayer ${\mathrm{mote}}_{2}$ and ${\mathrm{wse}}_{2}$},}\ }\href {\doibase 10.1103/PhysRevB.108.085117} {\bibfield  {journal} {\bibinfo  {journal} {Phys. Rev. B}\ }\textbf {\bibinfo {volume} {108}},\ \bibinfo {pages} {085117} (\bibinfo {year} {2023})}\BibitemShut {NoStop}%
\bibitem [{\citenamefont {Zhang}\ \emph {et~al.}(2024)\citenamefont {Zhang}, \citenamefont {Wang}, \citenamefont {Liu}, \citenamefont {Fan}, \citenamefont {Cao},\ and\ \citenamefont {Xiao}}]{zhangPolarizationdrivenBandTopology2024}%
  \BibitemOpen
  \bibfield  {author} {\bibinfo {author} {\bibfnamefont {Xiao-Wei}\ \bibnamefont {Zhang}}, \bibinfo {author} {\bibfnamefont {Chong}\ \bibnamefont {Wang}}, \bibinfo {author} {\bibfnamefont {Xiaoyu}\ \bibnamefont {Liu}}, \bibinfo {author} {\bibfnamefont {Yueyao}\ \bibnamefont {Fan}}, \bibinfo {author} {\bibfnamefont {Ting}\ \bibnamefont {Cao}}, \ and\ \bibinfo {author} {\bibfnamefont {Di}~\bibnamefont {Xiao}},\ }\bibfield  {title} {\enquote {\bibinfo {title} {Polarization-driven band topology evolution in twisted {{MoTe2}} and {{WSe2}}},}\ }\href {\doibase 10.1038/s41467-024-48511-x} {\bibfield  {journal} {\bibinfo  {journal} {Nature Communications}\ }\textbf {\bibinfo {volume} {15}},\ \bibinfo {pages} {4223} (\bibinfo {year} {2024})}\BibitemShut {NoStop}%
\bibitem [{\citenamefont {Xu}\ \emph {et~al.}(2024)\citenamefont {Xu}, \citenamefont {Mao}, \citenamefont {Zeng},\ and\ \citenamefont {Zhang}}]{xu2024multiple}%
  \BibitemOpen
  \bibfield  {author} {\bibinfo {author} {\bibfnamefont {Cheng}\ \bibnamefont {Xu}}, \bibinfo {author} {\bibfnamefont {Ning}\ \bibnamefont {Mao}}, \bibinfo {author} {\bibfnamefont {Tiansheng}\ \bibnamefont {Zeng}}, \ and\ \bibinfo {author} {\bibfnamefont {Yang}\ \bibnamefont {Zhang}},\ }\bibfield  {title} {\enquote {\bibinfo {title} {Multiple chern bands in twisted mote $ \_2 $ and possible non-abelian states},}\ }\href {https://doi.org/10.48550/arXiv.2403.17003} {\bibfield  {journal} {\bibinfo  {journal} {arXiv preprint arXiv:2403.17003}\ } (\bibinfo {year} {2024})}\BibitemShut {NoStop}%
\bibitem [{\citenamefont {Wang}\ \emph {et~al.}(2024)\citenamefont {Wang}, \citenamefont {Zhang}, \citenamefont {Liu}, \citenamefont {Wang}, \citenamefont {Cao},\ and\ \citenamefont {Xiao}}]{wang2024higher}%
  \BibitemOpen
  \bibfield  {author} {\bibinfo {author} {\bibfnamefont {Chong}\ \bibnamefont {Wang}}, \bibinfo {author} {\bibfnamefont {Xiao-Wei}\ \bibnamefont {Zhang}}, \bibinfo {author} {\bibfnamefont {Xiaoyu}\ \bibnamefont {Liu}}, \bibinfo {author} {\bibfnamefont {Jie}\ \bibnamefont {Wang}}, \bibinfo {author} {\bibfnamefont {Ting}\ \bibnamefont {Cao}}, \ and\ \bibinfo {author} {\bibfnamefont {Di}~\bibnamefont {Xiao}},\ }\bibfield  {title} {\enquote {\bibinfo {title} {Higher landau-level analogues and signatures of non-abelian states in twisted bilayer mote $ \_2$},}\ }\href {https://doi.org/10.48550/arXiv.2404.05697} {\bibfield  {journal} {\bibinfo  {journal} {arXiv preprint arXiv:2404.05697}\ } (\bibinfo {year} {2024})}\BibitemShut {NoStop}%
\bibitem [{\citenamefont {Ahn}\ \emph {et~al.}(2024)\citenamefont {Ahn}, \citenamefont {Lee}, \citenamefont {Yananose}, \citenamefont {Kim},\ and\ \citenamefont {Cho}}]{ahn2024first}%
  \BibitemOpen
  \bibfield  {author} {\bibinfo {author} {\bibfnamefont {Cheong-Eung}\ \bibnamefont {Ahn}}, \bibinfo {author} {\bibfnamefont {Wonjun}\ \bibnamefont {Lee}}, \bibinfo {author} {\bibfnamefont {Kunihiro}\ \bibnamefont {Yananose}}, \bibinfo {author} {\bibfnamefont {Youngwook}\ \bibnamefont {Kim}}, \ and\ \bibinfo {author} {\bibfnamefont {Gil~Young}\ \bibnamefont {Cho}},\ }\bibfield  {title} {\enquote {\bibinfo {title} {First landau level physics in second moir\'e band of $2.1^{\circ}$ twisted bilayer {${\rm MoTe}_2$}},}\ }\href {https://doi.org/10.48550/arXiv.2403.19155} {\bibfield  {journal} {\bibinfo  {journal} {arXiv preprint arXiv:2403.19155}\ } (\bibinfo {year} {2024})}\BibitemShut {NoStop}%
\bibitem [{\citenamefont {Chen}\ \emph {et~al.}(2024)\citenamefont {Chen}, \citenamefont {Luo}, \citenamefont {Zhu},\ and\ \citenamefont {Sheng}}]{chen2024robust}%
  \BibitemOpen
  \bibfield  {author} {\bibinfo {author} {\bibfnamefont {Feng}\ \bibnamefont {Chen}}, \bibinfo {author} {\bibfnamefont {Wei-Wei}\ \bibnamefont {Luo}}, \bibinfo {author} {\bibfnamefont {Wei}\ \bibnamefont {Zhu}}, \ and\ \bibinfo {author} {\bibfnamefont {DN}~\bibnamefont {Sheng}},\ }\bibfield  {title} {\enquote {\bibinfo {title} {Robust non-abelian even-denominator fractional chern insulator in twisted bilayer mote $ \_2$},}\ }\href {https://doi.org/10.48550/arXiv.2405.08386} {\bibfield  {journal} {\bibinfo  {journal} {arXiv preprint arXiv:2405.08386}\ } (\bibinfo {year} {2024})}\BibitemShut {NoStop}%
\bibitem [{\citenamefont {Abouelkomsan}\ \emph {et~al.}(2024)\citenamefont {Abouelkomsan}, \citenamefont {Reddy}, \citenamefont {Fu},\ and\ \citenamefont {Bergholtz}}]{abouelkomsan2024band}%
  \BibitemOpen
  \bibfield  {author} {\bibinfo {author} {\bibfnamefont {Ahmed}\ \bibnamefont {Abouelkomsan}}, \bibinfo {author} {\bibfnamefont {Aidan~P}\ \bibnamefont {Reddy}}, \bibinfo {author} {\bibfnamefont {Liang}\ \bibnamefont {Fu}}, \ and\ \bibinfo {author} {\bibfnamefont {Emil~J}\ \bibnamefont {Bergholtz}},\ }\bibfield  {title} {\enquote {\bibinfo {title} {Band mixing in the quantum anomalous hall regime of twisted semiconductor bilayers},}\ }\href {https://doi.org/10.1103/PhysRevB.109.L121107} {\bibfield  {journal} {\bibinfo  {journal} {Physical Review B}\ }\textbf {\bibinfo {volume} {109}},\ \bibinfo {pages} {L121107} (\bibinfo {year} {2024})}\BibitemShut {NoStop}%
\bibitem [{\citenamefont {Yu}\ \emph {et~al.}(2024)\citenamefont {Yu}, \citenamefont {Herzog-Arbeitman}, \citenamefont {Wang}, \citenamefont {Vafek}, \citenamefont {Bernevig},\ and\ \citenamefont {Regnault}}]{wubandmixing2024}%
  \BibitemOpen
  \bibfield  {author} {\bibinfo {author} {\bibfnamefont {Jiabin}\ \bibnamefont {Yu}}, \bibinfo {author} {\bibfnamefont {Jonah}\ \bibnamefont {Herzog-Arbeitman}}, \bibinfo {author} {\bibfnamefont {Minxuan}\ \bibnamefont {Wang}}, \bibinfo {author} {\bibfnamefont {Oskar}\ \bibnamefont {Vafek}}, \bibinfo {author} {\bibfnamefont {B.~Andrei}\ \bibnamefont {Bernevig}}, \ and\ \bibinfo {author} {\bibfnamefont {Nicolas}\ \bibnamefont {Regnault}},\ }\bibfield  {title} {\enquote {\bibinfo {title} {Fractional chern insulators versus nonmagnetic states in twisted bilayer ${\mathrm{mote}}_{2}$},}\ }\href {\doibase 10.1103/PhysRevB.109.045147} {\bibfield  {journal} {\bibinfo  {journal} {Phys. Rev. B}\ }\textbf {\bibinfo {volume} {109}},\ \bibinfo {pages} {045147} (\bibinfo {year} {2024})}\BibitemShut {NoStop}%
\bibitem [{\citenamefont {Cappelli}\ and\ \citenamefont {Randellini}(2015)}]{cappelliStabilityTopologicalInsulators2015}%
  \BibitemOpen
  \bibfield  {author} {\bibinfo {author} {\bibfnamefont {Andrea}\ \bibnamefont {Cappelli}}\ and\ \bibinfo {author} {\bibfnamefont {Enrico}\ \bibnamefont {Randellini}},\ }\bibfield  {title} {\enquote {\bibinfo {title} {Stability of topological insulators with non-{{Abelian}} edge excitations},}\ }\href {\doibase 10.1088/1751-8113/48/10/105404} {\bibfield  {journal} {\bibinfo  {journal} {Journal of Physics A: Mathematical and Theoretical}\ }\textbf {\bibinfo {volume} {48}},\ \bibinfo {pages} {105404} (\bibinfo {year} {2015})}\BibitemShut {NoStop}%
\bibitem [{\citenamefont {Park}\ \emph {et~al.}(2024)\citenamefont {Park}, \citenamefont {Cai}, \citenamefont {Anderson}, \citenamefont {Zhang}, \citenamefont {Liu}, \citenamefont {Holtzmann}, \citenamefont {Li}, \citenamefont {Wang}, \citenamefont {Hu}, \citenamefont {Zhao} \emph {et~al.}}]{park2024ferromagnetism}%
  \BibitemOpen
  \bibfield  {author} {\bibinfo {author} {\bibfnamefont {Heonjoon}\ \bibnamefont {Park}}, \bibinfo {author} {\bibfnamefont {Jiaqi}\ \bibnamefont {Cai}}, \bibinfo {author} {\bibfnamefont {Eric}\ \bibnamefont {Anderson}}, \bibinfo {author} {\bibfnamefont {Xiao-Wei}\ \bibnamefont {Zhang}}, \bibinfo {author} {\bibfnamefont {Xiaoyu}\ \bibnamefont {Liu}}, \bibinfo {author} {\bibfnamefont {William}\ \bibnamefont {Holtzmann}}, \bibinfo {author} {\bibfnamefont {Weijie}\ \bibnamefont {Li}}, \bibinfo {author} {\bibfnamefont {Chong}\ \bibnamefont {Wang}}, \bibinfo {author} {\bibfnamefont {Chaowei}\ \bibnamefont {Hu}}, \bibinfo {author} {\bibfnamefont {Yuzhou}\ \bibnamefont {Zhao}},  \emph {et~al.},\ }\bibfield  {title} {\enquote {\bibinfo {title} {Ferromagnetism and topology of the higher flat band in a fractional chern insulator},}\ }\href {https://arxiv.org/abs/2406.09591} {\bibfield  {journal} {\bibinfo  {journal} {arXiv preprint arXiv:2406.09591}\ } (\bibinfo {year} {2024})}\BibitemShut {NoStop}%
\bibitem [{\citenamefont {Haldane}(2018)}]{haldane_modular-invariant_2018}%
  \BibitemOpen
  \bibfield  {author} {\bibinfo {author} {\bibfnamefont {F.~D.~M.}\ \bibnamefont {Haldane}},\ }\bibfield  {title} {\enquote {\bibinfo {title} {A modular-invariant modified {Weierstrass} sigma-function as a building block for lowest-{Landau}-level wavefunctions on the torus},}\ }\href {\doibase 10.1063/1.5042618} {\bibfield  {journal} {\bibinfo  {journal} {Journal of Mathematical Physics}\ }\textbf {\bibinfo {volume} {59}},\ \bibinfo {pages} {071901} (\bibinfo {year} {2018})}\BibitemShut {NoStop}%
\bibitem [{\citenamefont {Wang}\ \emph {et~al.}(2021)\citenamefont {Wang}, \citenamefont {Cano}, \citenamefont {Millis}, \citenamefont {Liu},\ and\ \citenamefont {Yang}}]{wang_exact_2021}%
  \BibitemOpen
  \bibfield  {author} {\bibinfo {author} {\bibfnamefont {Jie}\ \bibnamefont {Wang}}, \bibinfo {author} {\bibfnamefont {Jennifer}\ \bibnamefont {Cano}}, \bibinfo {author} {\bibfnamefont {Andrew~J.}\ \bibnamefont {Millis}}, \bibinfo {author} {\bibfnamefont {Zhao}\ \bibnamefont {Liu}}, \ and\ \bibinfo {author} {\bibfnamefont {Bo}~\bibnamefont {Yang}},\ }\bibfield  {title} {\enquote {\bibinfo {title} {Exact {Landau} {Level} {Description} of {Geometry} and {Interaction} in a {Flatband}},}\ }\href {\doibase 10.1103/PhysRevLett.127.246403} {\bibfield  {journal} {\bibinfo  {journal} {Physical Review Letters}\ }\textbf {\bibinfo {volume} {127}},\ \bibinfo {pages} {246403} (\bibinfo {year} {2021})}\BibitemShut {NoStop}%
\bibitem [{\citenamefont {Gross}\ \emph {et~al.}(1991)\citenamefont {Gross}, \citenamefont {Runge},\ and\ \citenamefont {Heinonen}}]{gross1991many}%
  \BibitemOpen
  \bibfield  {author} {\bibinfo {author} {\bibfnamefont {E.K.U.}\ \bibnamefont {Gross}}, \bibinfo {author} {\bibfnamefont {E.}~\bibnamefont {Runge}}, \ and\ \bibinfo {author} {\bibfnamefont {O.}~\bibnamefont {Heinonen}},\ }\href@noop {} {\emph {\bibinfo {title} {Many-Particle Theory,}}}\ (\bibinfo  {publisher} {Taylor \& Francis},\ \bibinfo {year} {1991})\BibitemShut {NoStop}%
\end{thebibliography}%
\newpage
\begin{widetext}
    
		 \renewcommand{\theequation}{S\arabic{equation}}
		\setcounter{equation}{0}
		 \renewcommand{\thefigure}{S\arabic{figure}}
		\setcounter{figure}{0}
		 \renewcommand{\thetable}{S\arabic{table}}
		 \setcounter{table}{0}

\section{Supplemental Material}

\section{Landau Levels of opposite magnetic fields}

Consider two dimensional electrons subject to a magnetic field, the non-interaction Hamiltonian is given by \begin{equation}
    H = \frac{(\mathbf{p} - e \sigma_z \mathbf{A})^2}{2m}
\end{equation}
where $\sigma_z = \pm 1$ denotes the two opposite magnetic fields for the two opposite spins. We work in the symmetric gauge $\mathbf{A}_a = -B \epsilon_{ab} \mathbf{r}_b/2$. The canonical momentum along with its commutation relation are are defined as \begin{equation}
\begin{split}
        \boldsymbol{\pi}^{\sigma} = \mathbf{p} - e\sigma_z \mathbf{A}\\
        [\boldsymbol{\pi}^{\sigma}_{a},\boldsymbol{\pi}^{\sigma}_{b}] = i \sigma_z \epsilon_{ab} l^{-2}_B
\end{split}
\end{equation}
where $l_B = \sqrt{\hbar/eB}$ is the magnetic length. 
As standard in the quantum Hall problem, we define the guiding center coordinate and the cyclotron coordinate as \begin{equation}
    \begin{split}
        \mathbf{R}^{\sigma}_{a} = \mathbf{r}_{a} - l^2_B \epsilon_{ab} \boldsymbol{\pi}^{\sigma}_{b} = \mathbf{r}_a/2 - i \sigma_z \epsilon_{ab} \partial_b \\
        \bar{\mathbf{R}}^{\sigma}_a =  l^2_B \epsilon_{ab} \boldsymbol{\pi}^{\sigma}_{b} =   \mathbf{r}_a/2 + i \sigma_z \epsilon_{ab} \partial_b
    \end{split}
\end{equation}
They satisfy the following algebra (for equal spin)
\begin{equation}
    \begin{split}[\mathbf{R}^{\sigma}_a,\mathbf{R}^{\sigma}_b] = -i \sigma_z \epsilon_{ab} l^2_B  \\
    [\bar{\mathbf{R}}^{\sigma}_a,\bar{\mathbf{R}}^{\sigma}_b] = i \sigma_z \epsilon_{ab} l^2_B \\
    [\mathbf{R}^{\sigma}_a,\bar{\mathbf{R}}^{\sigma}_b] = 0
    \end{split}
\end{equation}
While for opposite spins, we have the following
\begin{equation}
    \begin{split}
        [\mathbf{R}^{\sigma}_{a},\mathbf{R}^{-\sigma}_{b}] = 0 \\
[\bar{\mathbf{R}}^{\sigma}_{a},\bar{\mathbf{R}}^{-\sigma}_{b}] = 0 \\
        [\mathbf{R}^{\sigma}_{a},\bar{\mathbf{R}}^{-\sigma}_{b}] = -i \sigma_z \epsilon_{ab} l^2_B
    \end{split}
\end{equation}
Magnetic translations are generated by the guiding center coordinates $t^{\sigma}(\mathbf{q}) = e^{i \mathbf{R}^{\sigma} \cdot \mathbf{q}}$ and they satisfy the following algebra \begin{equation}
\label{eq:magalgebra}
    t^{\sigma}(\mathbf{q}_1) t^{\sigma}(\mathbf{q}_2) = e^{i \sigma_z l^2_B \mathbf{q}_1 \cross \mathbf{q}_2} t^{\sigma}(\mathbf{q}_2) t^{\sigma}(\mathbf{q}_1)
\end{equation}
We define a magnetic unit cell that encloses $ 2\pi$ flux quanta. The magnetic lattice is then spanned by $\mathbf{a} = m \mathbf{a}_1 + n \mathbf{a}_2$ where $\mathbf{a}_1$ and $\mathbf{a}_2$ are two basis vectors chosen such that $|\mathbf{a}_1 \wedge \mathbf{a}_2| = 2 \pi l^2_B$ meaning that there is one flux quanta enclosed by the unit cell. This also gives rise to magnetic reciprocal lattice basis vectors $\mathbf{g}_1$ and $\mathbf{g}_2$ such that $\mathbf{a}_i \cdot \mathbf{g}_j = 2 \pi \delta_{ij}$. Therefore we can apply Bloch theorem to obtain magnetic Bloch wavefunctions which we denote by $\ket{n,\mathbf{k},\sigma}$ where $n$ is a LL index and $\mathbf{k}$ is Bloch momentum. In the symmetric gauge on a torus, these are given by the modified Weierstrass sigma functions \cite{haldane_modular-invariant_2018}. We refer the reader to the supplementary materials of  Ref. \cite{wang_exact_2021} and Ref. \cite{reddy2024non} for a detailed introduction. The Weierstrass sigma functions satisfy the boundary condition,
\begin{equation}
    t^{\sigma}(\mathbf{g}_i) \ket{n,\mathbf{k},\sigma} =  - e^{i \sigma_z l^2_B \mathbf{g}_i \wedge \mathbf{k}} \ket{n,\mathbf{k},\sigma}
\end{equation}

The action of the magnetic translation operator for generic momentum $t^{\sigma}(\mathbf{q})$ on the magnetic Bloch state is given by \begin{equation}
    t^{\sigma}(\mathbf{q}) \ket{n,\mathbf{k},\sigma} = e^{\frac{i}{2} l^2_B\sigma_z \mathbf{q} \wedge \mathbf{k}} \ket{n, \mathbf{k} + \mathbf{q},\sigma}
\end{equation}
if $\mathbf{q}$ is a reciprocal lattice basis vector then the state is transformed to itself giving rise to \begin{equation}
\begin{split}
        t^{\sigma}(\mathbf{g}_i) \ket{n,\mathbf{k},\sigma} =  e^{\frac{i}{2} \sigma_z l^2_B \mathbf{g}_i \wedge \mathbf{k}} \ket{n,\mathbf{k} + \mathbf{g}_i,\sigma} = - e^{i \sigma_z l^2_B \mathbf{g}_i \wedge \mathbf{k}} \ket{n,\mathbf{k},\sigma} \\
        \ket{n,\mathbf{k} + \mathbf{g}_i,\sigma} = - e^{\frac{i}{2}\sigma_z \mathbf{g}_i \wedge \mathbf{k}} \ket{n,\mathbf{k}\sigma} 
\end{split}
\end{equation}
Since, we are dealing with density-density interactions, we would like to evaluate the matrix elements of the density operator in the LL basis. We will utilize the transformation properties of $\ket{n,\mathbf{k},\sigma}$ under the magnetic translation operator and the fact that the relative coordinate operator can be expressed in terms of the raising and lowering operators of the LLs as follow \begin{equation}
\begin{split}
        \bar{R}^{\uparrow}_x = l_B (\bar{a}^{\dagger} + \bar{a})/(\sqrt{2}) \\
        \bar{R}^{\uparrow}_y = l_B (\bar{a}^{\dagger} -\bar{a})/(\sqrt{2} i ) \\ 
        \bar{R}^{\downarrow}_x = l_B (\bar{a}^{\dagger} + \bar{a})/(\sqrt{2}) \\
        \bar{R}^{\downarrow}_y =   l_B (\bar{a} - \bar{a}^{\dagger})/(\sqrt{2}i)
\end{split}
\end{equation}
where $\bar{a}^\dagger \ket{n,\mathbf{k},\sigma} = \sqrt{n+1} \ket{n+1,\mathbf{k},\sigma}$. The matrix elements of the density operator can be recast as 
\begin{equation}
\label{eq:denmatrixelements}
\braket{n_1,\mathbf{k}_1,\sigma}{ e^{i \mathbf{q} \cdot \mathbf{r}}|n_2,\mathbf{k}_2,\sigma} = \braket{n_1,\mathbf{k}_1,\sigma}{ e^{i \mathbf{q} \cdot \mathbf{\bar{R}_{\sigma}}}e^{i \mathbf{q} \cdot \mathbf{R_{\sigma}}}|n_2,\mathbf{k}_2,\sigma} = G^{\sigma}_{n_1 n_2}(\mathbf{q}) F^{\sigma}({\mathbf{k}_1,\mathbf{k}_2},\mathbf{q}) 
\end{equation} where 
\begin{align}
\label{eq:detailedformfactors}
       & G^{\uparrow}_{n_1,n_2}(\mathbf{q}) = e^{-l_B^2 |\mathbf{q}|^2/4} \sqrt{\dfrac{n_< !}{n_> !}} L_{n_<}^{|n_2-n_1|}\bigg(\dfrac{|\mathbf{q}|^2 l_B^2}{2}\bigg) \times  \begin{cases} 
      \big(\dfrac{ -l_B \bar{z}}{\sqrt{2}}\big)^{|n_2 - n_1|} & n_2 > n_1 \\
      \big(\dfrac{ l_B z}{\sqrt{2}}\big)^{|n_2 - n_1|} & n_1 \geq n_2
   \end{cases} \\
   & G^{\downarrow}_{n_1,n_2}(\mathbf{q}) = G^{\uparrow}_{n_2,n_1}(\mathbf{q}) \\
  & F^{\sigma}(\mathbf{k_1},\mathbf{k_2},\mathbf{q}) =  \eta_\mathbf{g} \delta_{\mathbf{q} ,\mathbf{k}_1- \mathbf{k}_2 - \mathbf{g}} e^{\frac{i}{2} \sigma_z l_B^2((\mathbf{k}_1+\mathbf{k}_2)\wedge \mathbf{g}) + \mathbf{k}_1 \wedge \mathbf{k}_2}
\end{align} 
where $n_> = {\rm max}(n_1,n_2)$ and $n_< = {\rm min}(n_1,n_2)$, $L_{a}^{b}(x)$ is the generalized Laguerre polynomial, $z = q_x + iq_y$ and  $\eta_{\mathbf{q}} = 1 $ if $\mathbf{q}/2$ is allowed reciprocal lattice vector and $-1$ otherwise. 
\section{Interacting Hamiltonian}
Consider normal ordered density-density interactions of the form \begin{equation}
\label{eq:Ham_int}
    \begin{split}
               \tilde{H} = \frac{1}{2S_M} \sum_{\mathbf{q} \sigma} :V(\mathbf{q}) \varrho_{\sigma}(\mathbf{q}) \varrho_{\sigma}(-\mathbf{q}):  +  U(\mathbf{q}): \varrho_{\sigma}(\mathbf{q}) \varrho_{-\sigma}(-\mathbf{q}):
    \end{split}
\end{equation}

where $\varrho_{\sigma}(\mathbf{q})$ is the \textit{unprojected} density operator for spin $\sigma$. In our magnetic bloch basis, we have $\varrho_{\sigma}(\mathbf{q}) = \sum_{n_1 n_2 \mathbf{k} \in \rm BZ} \langle n_1,\mathbf{k} + \mathbf{q}, \sigma |e^{i \mathbf{q} \cdot \mathbf{r}}| n_2, \mathbf{k},\sigma \rangle c^{\dagger}_{n_1 \sigma, \mathbf{k}+\mathbf{q}} c_{n_2 ,\sigma, \mathbf{k}}$. Using \eqref{eq:denmatrixelements}, we can directly the evaluate the matrix elements, $\langle n_1,\mathbf{k} + \mathbf{q}, \sigma |e^{i \mathbf{q} \cdot \mathbf{r}}| n_2, \mathbf{k},\sigma \rangle $. 

Projecting to the $n$-th Landau level therefore gives rise to 

\begin{equation}
\label{eq:Ham_int}
    \begin{split}
               \tilde{H} = \frac{1}{2S_M} \sum_{\mathbf{q} \sigma} :V_n(\mathbf{q}) \rho_{\sigma}(\mathbf{q}) \rho_{\sigma}(-\mathbf{q}):  +  U_n(\mathbf{q}): \rho_{\sigma}(\mathbf{q}) \rho_{-\sigma}(-\mathbf{q}):
    \end{split}
\end{equation}
with the \textit{projected} density operator $\rho_{\sigma}(\mathbf{q}) = \sum_{\mathbf{k} \in \rm BZ } F^{\sigma}(\mathbf{k}+\mathbf{q},\mathbf{k},\mathbf{q}) c^{\dagger}_{\sigma \mathbf{k} + \mathbf{q}} c_{\sigma \mathbf{k}}$ and the projected interactions given by \begin{equation}
\begin{split}
        V_n(\mathbf{q}) = V(q) G^{\sigma}_{n n}(\mathbf{q}) G^{\sigma}_{n n}(-\mathbf{q}) \\
        U_n(\mathbf{q}) = U(q) G^{\sigma}_{n n}(\mathbf{q}) G^{-\sigma}_{n n}(-\mathbf{q}) 
\end{split}
\end{equation}
with $F(\mathbf{k}_1,\mathbf{k}_2,\mathbf{q}$ and $G^{\sigma}_{n n}(\mathbf{q})$ are given in equation \eqref{eq:detailedformfactors}.
\section{Two body problem}
In this section, we solve the two body problem for two interacting particles closely following \cite{villadiego2024halperin}. We consider a two body interaction $V(\mathbf{r}) = V(\mathbf{r}_1 - \mathbf{r}_2) $ which is a function only of the relative distance. Let's define a guiding center of mass (COM) and guiding center relative coordinate as \begin{equation}
    \mathbf{R}^{\rm CM} = \dfrac{\mathbf{R}^{\sigma_1}_1 + \mathbf{R}^{\sigma_2}_2}{2}, \> \> \mathbf{R}^{\rm rel} = \mathbf{R}^{\sigma_1}_1 - \mathbf{R}^{\sigma_2}_2
\end{equation}
They satisfy the following commutation relation \begin{equation}
\begin{split}
        [\mathbf{R}^{\rm rel}_a,\mathbf{R}^{\rm rel}_b]=4[\mathbf{R}^{\rm CM}_a,\mathbf{R}^{\rm CM}_b] = -i  l^2_B  \epsilon_{ab} (\sigma_{1z} + \sigma_{2z})  \\
        [\mathbf{R}^{\rm CM}_a, \mathbf{R}^{\rm rel}_b] = -\frac{i}{2} l^2_B \epsilon_{ab} (\sigma_{1z} - \sigma_{2z})
\end{split}
\end{equation}
We see that depending on the spin of the particles either aligned or opposite, the physics of the problem can be different. We discuss the two cases below
\begin{itemize}
    \item $\sigma_{1z} = -\sigma_{2z}$
\end{itemize}

In this case, we have $[\mathbf{R}^{\rm rel}_a,\mathbf{R}^{\rm rel}_b]= [\mathbf{R}^{\rm CM}_a,\mathbf{R}^{\rm CM}_b] = 0$ and $[\mathbf{R}^{\rm CM}_a, \mathbf{R}^{\rm rel}_b] = -i l^2_B \epsilon_{ab} \sigma_{1z}$. Therefore the eigenstates can be labelled by the two coordinates $\mathbf{R}^{\rm rel}_{a}$ for $a = 1,2$. Let's denote them by $\ket{\mathbf{R}^{\rm rel}_{a},n_1,n_2, \sigma_1,\sigma_{2}}$. Since, the guiding COM coordinate doesn't commute with guiding relative coordinates, we can only label the states with one of them. We then evaluate the interaction matrix elements to get \begin{equation}
\label{eq:twobodyoppositematelem}
\begin{split}
       \langle \mathbf{R}^{\rm rel}_{a},n_1,n_2 ,\sigma_1,\sigma_{2} |\int \frac{d^2\mathbf{q}}{(2\pi)^2} U({\mathbf{q}}) e^{i \mathbf{q} \cdot (\mathbf{r}^{\sigma_1}_1 - \mathbf{r}^{\sigma_2}_2)}|\mathbf{R}^{\rm rel}_{a},m_1,m_2,\sigma_1,\sigma_{2} \rangle = \\
         \langle \mathbf{R}^{\rm rel}_{a},n_1,n_2 ,\sigma_1,\sigma_{2} |\int \frac{d^2\mathbf{q}}{(2\pi)^2} U({\mathbf{q}}) e^{i \mathbf{q} \cdot (\bar{\mathbf{R}}^{\sigma_1}_1 - \bar{\mathbf{R}}^{\sigma_2}_2+ \mathbf{R^{\rm rel}})}|\mathbf{R}^{\rm rel}_{a},m_1,m_2,\sigma_1,\sigma_{2} \rangle 
       \\ = \int \frac{d^2\mathbf{q}}{(2\pi)^2} U({\mathbf{q}}) G^{\sigma_1}_{n_1 m_1}(\mathbf{q}) G^{\sigma_2}_{n_2 m_2}(-\mathbf{q}) e^{i \mathbf{q} \cdot \mathbf{R}^{\rm rel}}
\end{split}
\end{equation}
If we focus on a single Landau level  denoted by $n$ and neglect mixing with the other Landau levels, then the two body energies are simply \begin{equation}
\label{eq:twobody_oppositespins}
    U_n(\mathbf{R}^{\rm rel}) = \int \frac{d^2\mathbf{q}}{(2\pi)^2} U({\mathbf{q}}) |G^{\sigma_1}_{n n}(\mathbf{q})|^2 e^{i \mathbf{q} \cdot \mathbf{R}^{\rm rel}} = \int \frac{d^2\mathbf{q}}{(2\pi)^2} U_n(\mathbf{q}) e^{i \mathbf{q} \cdot \mathbf{R}^{\rm rel}}
\end{equation}
\begin{itemize}
    \item $\sigma_{1z} = \sigma_{2z}$
\end{itemize}
    In this case $ [\mathbf{R}^{\rm CM}_a, \mathbf{R}^{\rm rel}_b] = 0$ and the problem reduces to the standard problem of Haldane pseudopotentials \cite{haldane1983fractional}. The Hamiltonian factorizes into a COM part and the relative coordinate part. The COM operators $\mathbf{R}^{\rm CM}$ describe a charged particle of $2e$ charge in magnetic field $[\mathbf{R}^{\rm CM}_a,\mathbf{R}^{\rm CM}_b] = -\frac{i}{2} l^2_B \sigma_{1z}$ and $\mathbf{R}^{\rm rel}$ describe a charged particle of $e/2$ charge $[\mathbf{R}^{\rm rel}_a,\mathbf{R}^{\rm rel}_b] = -2 i l^2_B \epsilon_{ab} \sigma_{1z}$. The two body interaction is a function only of the relative coordinates and the two-body energies (Haldane pseudopotentials) can be conveniently worked out in the symmetric gauge using the angular momemntum quantum number.  For our purposes later, we will list here the Haldane psuedopotentials of the Coulomb interactions $V(\mathbf{r}) = e^2/ \epsilon \mathbf{r}$ for the $n = 1 $ Landau level in addition to an important matrix element \cite{simon_landau_2013}
    \begin{equation}
    \label{eq:Haldane_1LL_supp}
        V^{m}_{1} = \dfrac{e^2}{\epsilon l_B} \dfrac{\Gamma(m+1/2)}{2 m !} \dfrac{(m-3/8)(m-11/8)}{(m-1/2)(m-3/2)}
    \end{equation}
        \begin{equation}
        \label{eq:identity}
        \langle n, m + n|V(\mathbf{r})|0,m \rangle  = \dfrac{e^2}{\epsilon l_B} \dfrac{\Gamma(m+1/2) \Gamma(n+1/2)}{2\sqrt{\pi m! n! (n+m)!}}
    \end{equation}
    where $\ket{n,m}$ is a state in the $n$th Landau level with \textit{relative} angular momentum $m$. Recall that the angular momentum of the $n$-th Landau level is $m -n$. So the Coulomb interaction has non-zero matrix elements only when the relative angular momentum is conserved.
    \section{Perturbative band mixing calculation}

Here, we consider the two body problem of holes in the $n = 1$ LL with mixing to the $n = 0$ LL discussed in the main text. In what follows, we use Coulomb interactions $V(\mathbf{q}) = \frac{2 \pi e^2}{\epsilon |\mathbf{q}|}$ We define the Landau level mixing parameter $\kappa = E_c/\Delta$ where $E_c = e^2/\epsilon a_M$ is the Coulomb energy and $\Delta$ is the gap between the two bands.  To first order in $\kappa$, the two body energies for the equal-spin interactions are simply equation \eqref{eq:Haldane_1LL_supp} while for the opposite-spin interaction interactions, they are given by equation \eqref{eq:twobody_oppositespins} with $ n = 1$.

The second order corrections to the equal-spin interactions from mixing with the $n = 0$ are given by \begin{equation}
   \delta V_{1}^m = - \dfrac{|\langle 1, m + 1|V(\mathbf{r})|0,m \rangle|^2}{\Delta^2} = - \kappa^2 \frac{4 \pi}{\sqrt{3}} \Bigg[\dfrac{\Gamma(m+1/2) \Gamma(3/2)}{2\sqrt{\pi m! (m+1)!}}\Bigg]^2
\end{equation}
where we have made use of the identity in equation \eqref{eq:identity}. In our minimal model, the second order process is simply the two holes with relative momentum $m$ in the $n = 1$ LL scattering to the $n = 0$ LL then scattering back to the $n = 1$ LL.

 To calculate the second order corrections to opposite-spin interactions, we express the opposite-spin interactions in the basis $\ket{\mathbf{R}^{\rm rel}_{a},n_1,n_2, \sigma_1,\sigma_{2}} $ by defining  $\ket{n_1,n_2, \sigma_1,\sigma_{2}}  \equiv F_{n_1 \sigma_1}^\dagger F_{n_2\sigma_2}^\dagger \ket{0}$ and considering instead a modified interacting Hamiltonian $H(\mathbf{R}^{\rm rel})$. This gives to 
 \begin{equation}
     H(\mathbf{R}^{\rm rel}) = \frac{1}{2} \sum_{n_1 n_2 n_3 n_4,\sigma -\sigma} U^{\sigma}_{n_1 n_2 n_3 n_4}(\mathbf{R}^{\rm rel}) F^{\dagger}_{n_1 \sigma} F^{\dagger}_{n_2 -\sigma} F_{n_3 -\sigma} F_{n_4 \sigma}
 \end{equation}
 where $U^{\sigma}_{n_1 n_2 n_3 n_4}(\mathbf{R}^{\rm rel})$ are the matrix elements of the opposite-spin interactions. Now consider an initial state of two holes in the $n = 1$. At second order in $\kappa$, there are three types of processes. two where a hole is scattered to the $ n = 0$ LL with the other hole fixed and a third process where both holes scatter to the $n = 1$ LL. We have then.
\begin{equation}
\begin{split}
        \delta U_1(\mathbf{R}^{\rm rel}) = -\dfrac{|\langle 0,1 ,\sigma,-\sigma|H(\mathbf{R}^{\rm rel})|,1,1,\sigma,-\sigma \rangle|^2}{\Delta^2} - \dfrac{|\langle 1,0 ,\sigma,-\sigma |H(\mathbf{R}^{\rm rel})|1,1,\sigma,-\sigma \rangle|^2}{\Delta^2} \\ - \dfrac{|\langle 0,0 ,\sigma,-\sigma |H(\mathbf{R}^{\rm rel})|1,1,\sigma,-\sigma \rangle|^2}{\Delta^2} = - \frac{1}{\Delta^2}\Bigg[\int \frac{d^2\mathbf{q}}{(2\pi)^2} V({\mathbf{q}}) G^{\sigma}_{0 1}(\mathbf{q}) G^{-\sigma}_{1 1}(-\mathbf{q})  e^{i \mathbf{q} \cdot \mathbf{R}^{\rm rel}}\Bigg]^2 \\ -\frac{1}{\Delta^2}\Bigg[\int \frac{d^2\mathbf{q}}{(2\pi)^2} V({\mathbf{q}}) G^{\sigma}_{1 1}(\mathbf{q}) G^{-\sigma}_{0 1}(-\mathbf{q})  e^{i \mathbf{q} \cdot \mathbf{R}^{\rm rel}}\Bigg]^2 - \frac{1}{2\Delta^2}\Bigg[\int \frac{d^2\mathbf{q}}{(2\pi)^2} V({\mathbf{q}})G^{\uparrow}_{0 1}(\mathbf{q}) G^{\downarrow}_{0 1}(-\mathbf{q})  e^{i \mathbf{q} \cdot \mathbf{R}^{\rm rel}}\Bigg]^2
\end{split}
\end{equation}
Putting everything together, we arrive at the expressions provided in the main text which we copy here. For equal-spin interactions, the full two body Coulomb energies are 
\begin{equation}
   \tilde{V}_{1}^{m} = \dfrac{1}{\Delta} \frac{1}{\sqrt{\sqrt{3}/4\pi}} \dfrac{\Gamma(m+1/2)}{2 m !} \dfrac{(m-3/8)(m-11/8)}{(m-1/2)(m-3/2)}  - \kappa^2 \frac{4 \pi}{\sqrt{3}}\Bigg[\dfrac{\Gamma(m+1/2) \Gamma(3/2)}{2\sqrt{\pi m! (m+1)!}}\Bigg]^2
\end{equation}
and for the opposite spin interactions interactions
\begin{equation}
\begin{split}
\tilde{U}_1(\mathbf{R}^{\rm rel}) =  \dfrac{1}{\Delta}\int \frac{d^2\mathbf{q}}{(2\pi)^2} V({\mathbf{q}}) |G^{\sigma}_{1 1}(\mathbf{q})|^2 e^{i \mathbf{q} \cdot \mathbf{R}^{\rm rel}} + \delta U_1(\mathbf{R}^{\rm rel})
\end{split}
\end{equation}
For $\mathbf{R}^{\rm Rel} = 0$, the expressions simplifies to \begin{equation}
   \tilde{U}_1(0) =  \kappa \frac{3 \pi}{4} \sqrt{\frac{1}{2 \sqrt{3}}} - \kappa^2 \frac{4 \pi^2}{16\sqrt{3}} 
\end{equation}
 \begin{figure}[t!]
    \centering
    \includegraphics[width = 0.6 \linewidth]{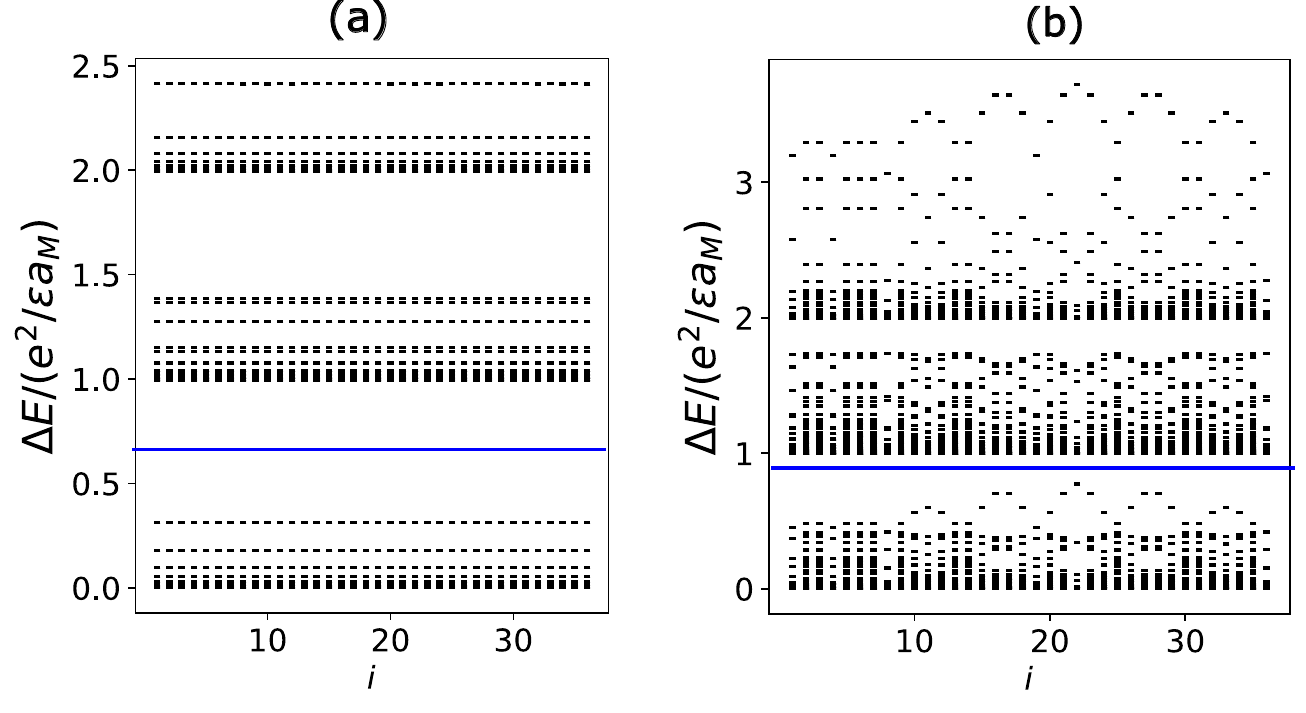}
    \caption{Coulomb two body energies for the two band model discussed in the main text which consists of the $n = 1$ LL and the $n = 0$ LL, separated by a gap $\Delta$ for (a) equal-spin interactions and (b) opposite-spin interactions evaluated at band mixing parameter $\kappa = 0.5$. The states below the blue line represent the first band which has interacting states of two holes in the $n = 1 $ LL.} 
     \label{fig:twobody_supp}
\end{figure}

\section{Numerical two-body problem}
In addition to the perturbative calculation, we numerically solve the two body problem of holes in the two band model discussed in the main text which consists of the $n = 1$ LL and the $n = 0$ LL, separated by a gap $\Delta$. In the absence of band mixing, the exact two-body spectrum for either equal-spin or opposite-spin interactions is divided into three \textit{bands}, the lowest band consists of interacting states of two holes in the $n = 1$ LL, the middle band has interacting states of a hole in the $n = 1$ and $ n= 0$ LL and the top band consits of interacting states of two holes in the $n = 0$ LL (Recall that in our problem, the $n = 1$ LL of holes is energetically lower than the $n = 0$ LL). 

We will focus on the lowest band since we want to calculate how the two body interaction in the $n = 1$ gets modified due to mixing. In the presence of band mixing, the energy gaps between the bands will change. We will focus on moderate band mixing values assuming the gap between the lowest band and the middle one doesn't change.

As a measure of the strength of long range Coulomb repulsion on finite systems, we define the interaction bandwidth $W_{\sigma_1\sigma_2}(\kappa) = E^{\sigma_1\sigma_2}_{\rm max}(\kappa) - E^{\sigma_1\sigma_2}_{\rm min}(\kappa)$ as the difference between the highest and lowest energies in the lowest band of the two-body spectrum for equal-spin interactions ($\sigma_1 = \sigma_2$) or opposite interactions ($\sigma_1 = - \sigma_2$) at a  specific value of $\kappa$. It's clear that $W_{\sigma_1\sigma_2}(\kappa)$ in the thermodynamic limit converges to the renormalized Haldane pseudopotential  $\tilde{V}^1_{1}$ of the $n = 1$ LL for equal- spin interactions and the renormalized $\tilde{U}_1(0)$ for opposite-spin interactions since $E^{\sigma_1 \sigma_2}_{\rm min} \rightarrow 0 $ as the Coulomb interaction vanishes at large distances.

In Fig. \ref{fig:twobody_supp}, we show the two-body spectrum at $\kappa = 0.5$. As evident, it consists of three bands and we focus only on the energies in the lowest band.  The plotted dots in Fig. 7 in the main text correspond to $W_{\sigma_1\sigma_2}(\kappa)$ calculated on a $6 \times 6$ finite system. At small $\kappa$, the numerically obtained values agrees well with the perturbative calculation.

\section{Interaction with neutralizing charge background}
Let's consider a generic band projected electron-electron interactions of the form 
\begin{equation}
\begin{split}
        H = \frac{1}{2A} \sum_{\mathbf{k}_1 \mathbf{k}_2 \mathbf{q} \sigma_1 \sigma_2} V_{\sigma_1 \sigma_2}(\mathbf{q}) F_{\sigma}(\mathbf{k}_1,\mathbf{q}) F_{\sigma}(\mathbf{k}_2,\mathbf{q}) c^{\dagger}_{\mathbf{k}_1 + \mathbf{q} \> \sigma_1} c^{\dagger}_{\mathbf{k}_2 - \mathbf{q} \> \sigma_2} c_{\mathbf{k}_2 \sigma_2} c_{\mathbf{k}_1 \sigma_1}
\end{split}
\end{equation}
where $V_{\sigma_1\sigma_2}(\mathbf{q})= \frac{2 \pi e^2}{\epsilon |\mathbf{q}| } e^{-|\mathbf{q}| d \delta_{\sigma_1, -\sigma_2}} $ is the interaction. $F_{\sigma}(\mathbf{k},\mathbf{q}) = \langle u_{\sigma}(\mathbf{k}+\mathbf{q})| u_{\sigma}(\mathbf{k}) \rangle $ is the form factors of the band wavefunctions and  $A$ is the area. We consider electrons of both spins interacting with the same uniform charge background. The Hamiltonian describing this interaction is given by \begin{equation}
    H_{\text{e-back}} = \frac{-e^2 N }{A} \sum_{\sigma} \int d^2\mathbf{r} \int d^2\mathbf{R} \frac{1}{|\mathbf{r}-\mathbf{R}|} \psi_{\sigma}^\dagger(\mathbf{r}) \psi_{\sigma}(\mathbf{r})
\end{equation}
Fourier transforming the above Hamiltonian, we have \begin{equation}
\begin{split}
        H_{\text{e-back}} = \frac{-e^2 N}{A} \sum_{\sigma}\sum_{\mathbf{k}} V_{\sigma \sigma}(0) \psi^\dagger_{\sigma}(\mathbf{k}) \psi_{\sigma}(\mathbf{k}) \\
         = \frac{- N }{A} V(0) (N_{\up} + N_{\downarrow}) = \frac{- N^2 V(0)}{A}
\end{split}
\end{equation}
where $V(0) = \text{lim}_{\mathbf{q} \rightarrow 0}\frac{2 \pi e^2}{\epsilon |\mathbf{q}|}$. In addition, the uniform background charge interacts electrostatically with itself, \begin{equation}
    \begin{split}
    H_{\text{back}} = \frac{e^2 N^2}{2 A^2} \int d^2 \mathbf{R} \int d^2 \mathbf{R}' \frac{1}{|\mathbf{R}-\mathbf{R}'|} 
    = \frac{N^2 V(0)}{2 A} 
    \end{split}
\end{equation}

Now, we take the limit $\mathbf{q} \rightarrow 0$ in the electron-electron Hamiltonian, we end up with two contributions, $H_{\text{lim}_{\mathbf{q} \rightarrow 0}} = H_1 + H_2$ \begin{equation}
\begin{split}
       H_1 =  \frac{1}{2A} 
 \sum_{\sigma} \text{lim}_{\mathbf{q} \rightarrow 0} V_{\sigma\sigma}(\mathbf{q}) N_{\sigma} (N_{\sigma} - 1) \\
 H_2 = \frac{1}{2A} 
 \sum_{\sigma_1 \neq \sigma_2} \text{lim}_{\mathbf{q} \rightarrow 0}  V_{\sigma_1 \sigma_2}(\mathbf{q}) N_{\sigma_1} N_{\sigma_2}
\end{split}
\end{equation}
Using $N_{\uparrow} N_{\downarrow} = N^2/4 -  S_z^2$ and $N_{\uparrow}^2 + N_{\downarrow}^2 = N^2/2 +  2 S_z^2$ where $S_z = (N_{\up} - N_{\downarrow})/2$, we end up with 
\begin{equation}
    \begin{split}
   H_1 + H_2 =  \frac{1}{2A} \text{lim}_{\mathbf{q} \rightarrow 0} (V_{\uparrow \uparrow}(\mathbf{q}) + V_{\uparrow \downarrow}(\mathbf{q})) \frac{N^2}{2} \\ + \frac{1}{A} \text{lim}_{\mathbf{q} \rightarrow 0} (V_{\uparrow \uparrow}(\mathbf{q}) - V_{\uparrow \downarrow}(\mathbf{q})) S^2_z  - \frac{1}{2A} V(0) N 
    \end{split}
\end{equation}
We observe that $\text{lim}_{\mathbf{q} \rightarrow 0} (V_{\uparrow \uparrow}(\mathbf{q}) - V_{\uparrow \downarrow}(\mathbf{q})) = \frac{2 \pi e^2}{\epsilon} d $ and $\text{lim}_{\mathbf{q} \rightarrow 0} (V_{\uparrow \uparrow}(\mathbf{q}) + V_{\uparrow \downarrow}(\mathbf{q})) = 2 V(0) $ so we get \begin{equation}
    H_{\text{lim}_{\mathbf{q} \rightarrow 0}} = \frac{N^2 V(0)}{2 A} - \frac{N V(0)}{2 A} +  \frac{2 \pi e^2}{\epsilon} \frac{d S^2_z}{A}
\end{equation}
We see that the first term is cancelled with $H_{\text{back}}$ and $H_{\text{e-back}}$. The second term can be neglected in the thermodynamic limit \cite{gross1991many}. We are therefore left with the third term which needs to be taken into account when comparing energies in different spin sectors. 

         \end{widetext}

\end{document}